\documentclass[11pt]{article}
\pdfoutput=1
\usepackage[a4paper,margin=3cm,bmargin=4.5cm]{geometry}
\usepackage{graphicx}
\usepackage{xcolor,amssymb,amsmath}
\usepackage{slashed}
\usepackage{hyperref}
\usepackage{booktabs}
\usepackage{braket}
\usepackage{subcaption}
\usepackage{slashed}
\newcommand{\Slash}[1]{{\ooalign{\hfil#1\hfil\crcr\raise.167ex\hbox{/}}}}
\usepackage[separate-uncertainty=true]{siunitx}
\usepackage{cite}
\usepackage{cancel}
\usepackage[utf8]{inputenc}
\usepackage{floatpag}

\usepackage[compat=1.1.0]{tikz-feynman}
\tikzfeynmanset{
  every edge={thick},
  warn luatex=false, 
}

\makeatletter
\def\tikzfeynman@luatex@required@path{}
\def\tikzfeynman@luatex@required@key{}
\makeatother

\hypersetup{
  colorlinks = true,
  linkcolor = blue,
  citecolor = blue,
}

\let\tilde\widetilde

\def\beq#1\eeq{\begin{align}#1\end{align}}

\newcommand\w[1]{_{\mathrm{#1}}}

\DeclareMathOperator{\Order}{\mathcal{O}}

\DeclareMathOperator{\Br}{\mathrm{Br}}

\renewcommand\Re{\mathop{\mathrm{Re}}}

\newcommand{\eg}{{\em e.g.}}
\newcommand{\ie}{{\em i.e.}}
\newcommand\TO{\text{--}}

\newcommand\unit[1]{\,\mathrm{#1}}
\newcommand\GeV{\unit{GeV}}
\newcommand\TeV{\unit{TeV}}

\newcommand\ifb{\unit{fb^{-1}}}

\newcommand\mET{\cancel{E}_{\mathrm T}}

\newcommand\amu[1][\relax]{\ifx#1\relax{a_\mu}\else{a_\mu^{\mathrm{#1}}}\fi}

\newcommand\smuL{\tilde\mu\w L}

\newcommand\neut  [1][\relax]{{\tilde\chi^0_{#1}}}
\newcommand\charP [1][\relax]{{\tilde\chi^+_{#1}}}
\newcommand\charM [1][\relax]{{\tilde\chi^-_{#1}}}
\newcommand\charPM[1][\relax]{{\tilde\chi^\pm_{#1}}}
\newcommand{\gL}{g\w L}
\newcommand{\gR}{g\w R}

\newcommand\package[2][\relax]{\texttt{#2}\ifx#1\relax\relax\else~\texttt{#1}\fi}

\newcommand\UL{_{\text{UL}}}
\newcommand\ULX{_{\text{UL;$X$}}}
\newcommand\ULorig{_{\text{UL;original}}}

\newcommand\pmat[1]{\begin{pmatrix}#1\end{pmatrix}}

\newcommand{\figref}[1]{Fig.~\ref{fig:#1}}
\newcommand{\figsref}[2]{Figs.~\ref{fig:#1} and \ref{fig:#2}}
\newcommand{\Figref}[1]{Figure~\ref{fig:#1}}

\newcommand\UPDATED[1]{\textbf{\sffamily[#1]}}

\begin{document}

\begin{titlepage}
\setcounter{page}{0} 
\begin{flushright}
UT--20--01\\
IPMU20--0009\\
KEK--TH--2188
\end{flushright}
\vskip 1.5cm
\begin{center}
  {\Large \bf Muon $\boldsymbol{g-2}$ vs LHC Run 2 in Supersymmetric Models
  }
\vskip 1.5cm
{
  Motoi Endo,$^{(a,b,c)}$
  Koichi Hamaguchi,$^{(c,d)}$
  Sho Iwamoto,$^{(e,f,g)}$
  and
  Teppei Kitahara$^{(h,i,j)}$
}

\vspace{1.5em}

\begingroup\small\itshape
\begin{tabbing}
$^{(a)}$ \!\!\= KEK Theory Center, IPNS, KEK, Tsukuba, Ibaraki 305--0801, Japan
\\[0.3em]
$^{(b)}$ \> The Graduate University of Advanced Studies (Sokendai), Tsukuba, Ibaraki 305--0801, Japan
\\[0.3em]
$^{(c)}$ \> Kavli IPMU (WPI), UTIAS, The University of Tokyo, Kashiwa, Chiba 277--8583, Japan
\\[0.3em]
$^{(d)}$ \> Department of Physics, The University of Tokyo, Bunkyo-ku, Tokyo 113--0033, Japan
\\[0.3em]
$^{(e)}$ \> Universit\a`a degli Studi di Padova, Via Marzolo 8, Padua I-35131, Italy
\\[0.3em]
$^{(f)}$ \> INFN, Sezione di Padova, Via Marzolo 8, Padua I-35131, Italy
\\[0.3em]
$^{(g)}$ \> ELTE E\a"otv\a"os Lor\a'and University, P\a'azm\a'any P\a'eter s\a'et\a'any 1/A, Budapest H-1117, Hungary\\[0.3em]
$^{(h)}$ \> Physics Department, Technion---Israel Institute of Technology, Haifa 3200003, Israel
\\[0.3em]
$^{(i)}$ \> Institute for Advanced Research, Nagoya University, Nagoya 464--8601, Japan
\\[0.3em]
$^{(j)}$ \> Kobayashi-Maskawa Institute for the Origin of Particles and the Universe, \\
         \> Nagoya University,  Nagoya 464--8602, Japan
\end{tabbing}
\endgroup

\vspace{1cm}

\abstract{
\noindent 
Supersymmetric models with sub-TeV charginos and sleptons have been a candidate for the origin of the long-standing discrepancy in the muon anomalous magnetic moment ($g-2$).
By gathering all the available LHC Run~2 results, we investigate the latest LHC constraints on models that explain the anomaly by their chargino contribution to the muon $g-2$.
It is shown that the parameter regions where sleptons are lighter than charginos are strongly disfavored.
In contrast, we find that the models with $m_{\tilde{\mu}_{\rm L}}\gtrsim m_{\tilde{\chi}^{\pm}_1}$ are still widely allowed, where the lighter chargino dominantly decays into a $W$-boson and a neutralino.
}

\vspace{2em}
\end{center}
\end{titlepage}

\renewcommand{\thefootnote}{\#\arabic{footnote}}
\setcounter{footnote}{0}

\hrule
\tableofcontents
\vskip .2in
\hrule
\vskip .4in

\section{Introduction}
\label{sec:introduction}
There is a long-standing discrepancy between the theory and experimental value in the anomalous magnetic dipole moment of the muon (muon $g-2$),
\begin{equation}
 a_{\mu} \equiv \frac{g_{\mu}-2}{2}\,,
\end{equation}
where $g_{\mu}$ is the magnetic moment of the muon. 
The Standard Model (SM) value is composed of the QED\cite{Aoyama:2012wk,Aoyama:2017uqe}, electroweak\cite{Knecht:2002hr,Czarnecki:2002nt,Gnendiger:2013pva,Ishikawa:2018rlv}, 
hadronic vacuum polarization\cite{Keshavarzi:2019abf,Davier:2019can}, and hadronic light-by-light \cite{Prades:2009tw,Blum:2019ugy} contributions.\footnote{%
 It was reported \cite{Volkov:2019phy} that an independent estimation of the purely photonic five-loop contribution to $a_e$ became inconsistent with the result of Ref.~\cite{Aoyama:2017uqe}. Besides, see, \eg, Ref.~\cite{Keshavarzi:2019abf} for recent progresses on the hadronic light-by-light contribution.
}
The latest results are
\begin{equation}
 a_{\mu}^{\rm SM} = 
\begin{cases}
\left( 11\,659\,181.08 \pm 3.78 \right) \times 10^{-10} & \text{\cite{Keshavarzi:2019abf}},\\
\left( 11\,659\,183.0  \pm 4.8  \right) \times 10^{-10} & \text{\cite{Davier:2019can}},
\end{cases}
\end{equation}
depending especially on estimation of the hadronic vacuum polarization.
On the other hand, the result of the E821 experiment at Brookhaven is  \cite{Bennett:2002jb,Bennett:2004pv,Bennett:2006fi}\footnote{%
The value of $a_{\mu}^{\rm BNL}$ we quote is calculated with the latest value of the muon-to-proton magnetic ratio,
$\mu_{\mu}/\mu_p = -3.183\,345\,142(71)$, taken from the 2018 CODATA recommended values \cite{CODATA2018}.}
\begin{equation}
 a_{\mu}^{\rm BNL}  = \left(11\,659\,208.9 \pm 5.4_{\rm stat} \pm 3.3_{\rm sys} \right) \times 10^{-10}\,.
\label{eq:amuBNL}
\end{equation}
These values correspond to $3.8$ and $3.3\,\sigma$ level discrepancies, respectively:
\begin{equation}
 \Delta a_{\mu} \equiv a_{\mu}^{{\rm exp}} - a_{\mu}^{\rm SM} =
\begin{cases}
\left( 27.8 \pm 7.4\right) \times 10^{-10} & \text{\cite{Keshavarzi:2019abf}},\\
\left( 26.1 \pm 7.9\right) \times 10^{-10} & \text{\cite{Davier:2019can}}.
\end{cases}
\end{equation}
A new measurement of $a_\mu$ is underway at Fermilab and their first result is expected to be announced in the near future \cite{Grange:2015fou,Keshavarzi:2019bjn}. 
Besides, an independent experiment with new techniques is in preparation at J-PARC 
\cite{Mibe:2011zz, Abe:2019thb}.

The discrepancy is as large as the SM electroweak contribution, $a_\mu(\text{EW})= (15.4\pm 0.1)\times 10^{-10}$. 
This implies that physics beyond the SM (BSM) exists in a scale around or lower than the scale of order $100\GeV$--$1\TeV$.
Among various BSM solutions to the muon $g-2$ discrepancy, low energy supersymmetry (SUSY) is still one of the most attractive models. 
The SUSY contributions to the muon $g-2$ ($\amu[SUSY]$) can naturally saturate the discrepancy, 
when electroweakinos and sleptons have masses of $\Order(100)\GeV$, while keeping its advantages such as the stability of the electroweak scale, the gauge coupling unification, and the existence of the dark matter candidate as the lightest SUSY particle (LSP).

In this paper, we revisit the SUSY explanation for the muon $g-2$ anomaly.
The electroweakinos and sleptons of $\Order(100)\GeV$ are targets of LHC searches.
We aim to provide a direct test of the SUSY scenario at the LHC.
We consider the minimal model setup, \ie, it is assumed that the SUSY particles relevant for the muon $g-2$ are light, while the irrelevant ones are decoupled. 
We investigate whether those particles are already excluded by/survive the latest LHC constraints.

In particular, we focus on the parameter regions where the chargino contribution to the muon $g-2$ is dominant.
As it is likely to be larger than the other contributions,  it is important to clarify their viability after the LHC Run~2. 
Such a setup has been examined in our previous study \cite{Endo:2013bba} based on the early results from the LHC Run~1\cite{Aad:2012pxa,Aad:2012hba,ATLAS:2012uks}.
The LHC sensitivities have been improved in various aspects afterwards. 
In Ref.~\cite{Endo:2013bba}, pair-productions of electroweakinos decaying into the SM bosons were too weak to constrain the SUSY models considered.
However, studies of these channels as well as those for the slepton pair-production have progressed significantly. 
We will show that they become newly relevant for the scenario.
Moreover, the sensitivity of the three lepton channel, which gave the leading constraints in Ref.~\cite{Endo:2013bba}, is improved as well.
The list of the LHC Run~2 results used/checked in our study is summarized in Appendix~\ref{app:LHC_analysis_list}. 
We will also see that a large parameter region still survives the LHC constraints.\footnote{%
  One may consider several scenarios that explain $\Delta\amu$ but are very elusive at collider experiments.
  For example, models with $m_{\tilde\mu\w L}$ and $m_{\tilde\mu\w R}$ being $\Order(100)\GeV$ and $\mu\gg1\TeV$ may provide the BLR contribution, $\amu[BLR]$ of Eq.~\eqref{eq:BLR}, large enough to explain the anomaly, but it is very challenging to search for such models at colliders if $M_1\simeq m_{\tilde\mu\w L}\simeq m_{\tilde\mu\w R}$~\cite{Endo:2013lva,Aad:2019qnd}.
}

For recent studies of the muon $g-2$ motivated SUSY models based on the LHC Run~2 results, see, \eg, Refs.~\cite{Zhu:2016ncq,Choudhury:2017fuu,Yanagida:2017dao,Endo:2017zrj,Hagiwara:2017lse,Chakraborti:2017vxz,Choudhury:2017acn,Ajaib:2017zba,Belyaev:2018vkl,Bhattacharyya:2018inr,Abel:2018ekz,Cao:2018rix,Dutta:2018fge,Cox:2018vsv,Tran:2018kxv,Ibe:2019jbx,Badziak:2019gaf,Abdughani:2019wai,Yanagida:2020jzy}.

This paper is organized as follows. 
We describe SUSY parameter regions focused in this paper in Sec.~\ref{sec:setup} and show our main results in Fig.~\ref{fig:summary}.
The detailed analyses of the LHC bounds are given in Sec.~\ref{sec:LHC}, and 
Sec.~\ref{sec:conclusion} is devoted to conclusions and discussion. In Appendix~\ref{app:Auxiliary}, we provide extra materials and additional discussions of our study, and in Appendix~\ref{app:xs}, the relations of the neutralino-chargino production cross sections are discussed.
Appendix~\ref{app:LHC_analysis_list} summarizes the list of the LHC Run~2 results used/checked in our analysis.

\section{Setup}
\label{sec:setup}

In SUSY models, radiative corrections to the muon $g-2$ can be amplified 
when $\tan \beta$ is sizable and at least 
three SUSY multiplets are as light as $\mathcal{O}(100)\GeV$, where $\tan\beta=v_u/v_d$ is the ratio of the vacuum expectation values of the up- and down-type Higgs.
They are classified into four types: WHL, BHL, BHR, and BLR.
Each of the names describes the SUSY particles yielding the contribution; B, W, H, L, and R stand for bino, wino, higgsino, left-handed and right-handed sleptons, respectively.
They are given under the mass-insertion approximation by \cite{Moroi:1995yh}\footnote{In the numerical calculation, we do not use these approximations but use the package \package[1.5.0]{GM2Calc}~\cite{Athron:2015rva}. See discussion below.}
\begin{align}
  \amu[WHL]
    &=\frac{\alpha_2}{4\pi} \frac{m_{\mu}^2}{M_2\mu} \tan\beta\cdot
    f_C\left(\frac{M_2^2}{m_{\tilde{\nu}_{\mu}}^2}, \frac{\mu^2}{m_{\tilde{\nu}_{\mu}}^2} \right) -\frac{\alpha_2}{8\pi} \frac{m_{\mu}^2}{M_2\mu} \tan\beta\cdot
    f_N\left(\frac{M_2^2}{m_{\tilde{\mu}\w L}^2}, \frac{\mu^2}{m_{\tilde{\mu}\w L}^2} \right)\,,
    \label{eq:WHL} \\
  \amu[BHL]
  &= \frac{\alpha_Y}{8\pi} \frac{m_{\mu}^2}{M_1\mu} \tan\beta\cdot
    f_N\left(\frac{M_1^2}{m_{\tilde{\mu}\w L}^2}, \frac{\mu^2}{m_{\tilde{\mu}\w L}^2} \right)\,,
    \label{eq:BHL} \\
  \amu[BHR]
  &= - \frac{\alpha_Y}{4\pi} \frac{m_{\mu}^2}{M_1\mu} \tan\beta\cdot
    f_N\left(\frac{M_1^2}{m_{\tilde{\mu}\w R}^2}, \frac{\mu^2}{m_{\tilde{\mu}\w R}^2} \right)\,,
    \label{eq:BHR} \\
  \amu[BLR]
  &= \frac{\alpha_Y}{4\pi} \frac{m_{\mu}^2M_1\mu}{m_{\tilde{\mu}\w L}^2m_{\tilde{\mu}\w R}^2}
    \tan\beta\cdot
    f_N\left(\frac{m_{\tilde{\mu}\w L}^2}{M_1^2}, \frac{m_{\tilde{\mu}\w R}^2}{M_1^2} \right)\,,
    \label{eq:BLR} 
\end{align}
where $M_1$ ($M_2$) is the bino (wino) soft-mass parameter, $\mu$ is the higgsino mass parameter, and $m_{\tilde{\mu}\w {L/R}}$ and $m_{\tilde{\nu}_{\mu}}$ are the masses of the left/right-handed smuon and the muon sneutrino, respectively.
Note that $a_\mu^{\mathrm{WHL}}$ is the sum of the charged- and neutral-wino contributions (cf.~Refs.~\cite{Endo:2013bba,Endo:2017zrj}).
The loop functions are given by
\begin{align}
    \label{eq:loop-aprox}
    f_C(x,y)
    &=xy\left[
      \frac{5-3(x+y)+xy}{(x-1)^2(y-1)^2} - \frac{2\ln x}{(x-y)(x-1)^3}+\frac{2\ln y}{(x-y)(y-1)^3}
      \right]\,,
      \\
    f_N(x,y)
    &= xy\left[
      \frac{-3+x+y+xy}{(x-1)^2(y-1)^2} + \frac{2x\ln x}{(x-y)(x-1)^3}-\frac{2y\ln y}{(x-y)(y-1)^3}
      \right]\,,
\end{align}
which satisfy $0\le f_{C/N}(x,y)\le 1$, $f_C(1,1) = 1/2$ and $f_N(1,1) = 1/6$.
When all the SUSY particle masses are the same size, the chargino contribution $\amu[WHL]$ coming from the wino, higgsino, and left-handed smuon is larger by an order of magnitude than the others.

This work focuses on the scenarios in which the chargino contribution dominates the SUSY contributions to the muon $g-2$. 
Among SUSY particles, neutralinos $\neut[i]$, charginos $\charPM[j]$, and left-handed sleptons $\tilde l\w L$, $\tilde\nu$ are left within the LHC reach, \ie, with masses of $\lesssim1\TeV$, while the other SUSY particles are decoupled.
Such a setup is minimal to realize sizable chargino contributions, and thus, provides a direct test of the scenario at the LHC.
More specifically, our set-up is summarized as follows.
\begin{itemize}
\item All the colored SUSY particles (gluino and squarks) are decoupled.
This assumption makes the LHC constraints more conservative, while their contribution to the muon $g-2$ are negligibly small even if they are light because it arises at the two-loop level.
Heavy colored SUSY particles are motivated by the mass of the SM-like Higgs boson as well as by LHC constraints.
\item The heavy Higgs bosons, whose contributions to $\amu[SUSY]$ are also negligible, are decoupled as well.\footnote{%
Note that the heavy Higgs bosons are already strongly constrained by $p p \to H/A \to \tau \tau$ searches when $\tan \beta $ is sizable.
For instance, for $\tan\beta=40$, the ATLAS and CMS collaborations set a bound of
$M_A \gtrsim 1.5\TeV$ 
by $\sqrt{s}=13\TeV$ and $\int\! dt \mathcal{L}=36\ifb$ \cite{Aaboud:2017sjh, Sirunyan:2018zut}.
In such a case, the heavy Higgs bosons are well decoupled in the neutral CP-even Higgs sector, and we have checked that branching ratios of the electroweakinos, which are relevant parameters in our analysis, are stable against the choice of the masses.
}
\item The right-handed sleptons $\tilde l\w R$ are heavy.
Then, the BHR and BLR contributions to the muon $g-2$ are suppressed, and hence, our analysis is simplified. We also neglect the scalar trilinear terms $(A_e)_{ij}$ for simplicity.
\item The soft masses of the left-handed sleptons are flavor universal and diagonal.
\end{itemize}
Then, the following five model parameters are left relevant,
\begin{equation}
    M_1, M_2, \mu, m^2\w L, \tan\beta,
    \label{eq:parameters}
\end{equation}
where $m\w L$ represents the universal soft mass for the left-handed sleptons. 
Although the bino mass is irrelevant for $\amu[WHL]$, it is left small such that the LSP is the bino-like lightest neutralino.
In the present analysis, we consider the following four subspace of the parameters:\footnote{%
The relation $M_1 =M_2/2$ in (A) and (B) is inspired by the GUT relation.
This setup is similar to the ones used in the previous work~\cite{Endo:2013bba}, in which we assumed that $M_1:M_2:M_3=1:2:6$.
See also the discussion in Sec.~\ref{sec:conclusion}.
}
\begin{equation}
\begin{split}
 \text{(A)}\;
 & M_1=\frac{1}{2}M_2, \;\mu=M_2,\; \tan\beta=40,
 \\
 \text{(B)}\;
 & M_1=\frac{1}{2}M_2, \;\mu=2M_2,\; \tan\beta=40,
 \\
 \text{(C)}\;
 & m_{\tilde{\chi}^0_1} = 100\,\textrm{GeV},\; \mu=M_2,\; \tan\beta=40,
 \\
 \text{(D)}\;
 & m_{\tilde{\chi}^0_1} = 100\,\textrm{GeV},\; \mu=2M_2, \;\tan\beta=40.
\end{split}
\label{eq:parameterspace}
\end{equation}
In the cases (C) and (D), $M_1$ is chosen to have the same sign as $M_2$.
Two free parameters are left in each subspace.

In the following analysis, the SUSY mass spectra and mixing matrices are calculated at the tree level.
Decay rates of SUSY particles are calculated by \package[1.5a]{SDECAY} \cite{Muhlleitner:2003vg,Djouadi:2006bz}.
The SM parameters are taken from Ref.~\cite{Tanabashi:2018oca} except for the Higgs boson mass, which we fix to be $125.0\GeV$.

The values of $\amu[SUSY]$ are calculated by \package[1.5.0]{GM2Calc} \cite{Athron:2015rva} with the $\tan\beta$ resummation.
Although it can evaluate the relevant contributions to $\amu[SUSY]$ at the two-loop level~\cite{Fargnoli:2013zda,Fargnoli:2013zia}, we do not include them and use the one-loop result so that $\amu[SUSY]$ is independent of the masses of the decoupled SUSY particles.
Since we assume that $\tilde\mu\w R$ is decoupled and $M_2\leq\mu$,  $\amu[SUSY]$ is dominated by $\amu[WHL]$ with a subleading effect from $\amu[BHL]$:
\begin{equation}
 \amu[SUSY] \simeq \amu[WHL] + \amu[BHL]\,.
\end{equation}
Our estimation is subject to uncertainties coming from radiative corrections to the masses and mixings,
higher order contributions to $\amu[SUSY]$, and
the right-handed slepton mass parameter $m\w R$, which we take $m\w R = 3\TeV$ in the numerical calculation.

\begin{figure}[p]
 \centering
 \renewcommand\thesubfigure{\Alph{subfigure}}
  \begin{subfigure}[b]{0.49\textwidth}
 \includegraphics[width=\textwidth]{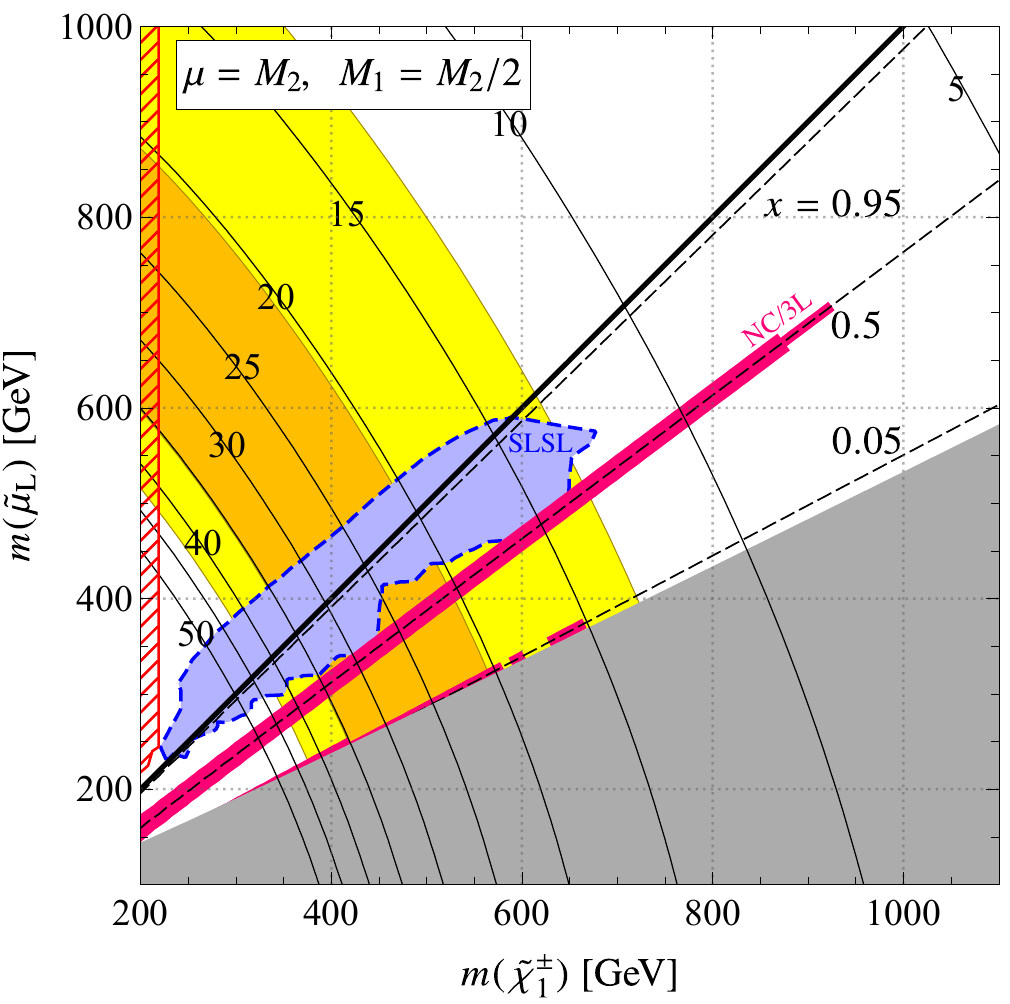}
\caption{$\mu = M_2$, $M_1 = M_2/2$}\label{fig:1gut}
 \vspace{.2cm}
 \end{subfigure}
   \begin{subfigure}[b]{0.49\textwidth}
 \includegraphics[width=\textwidth]{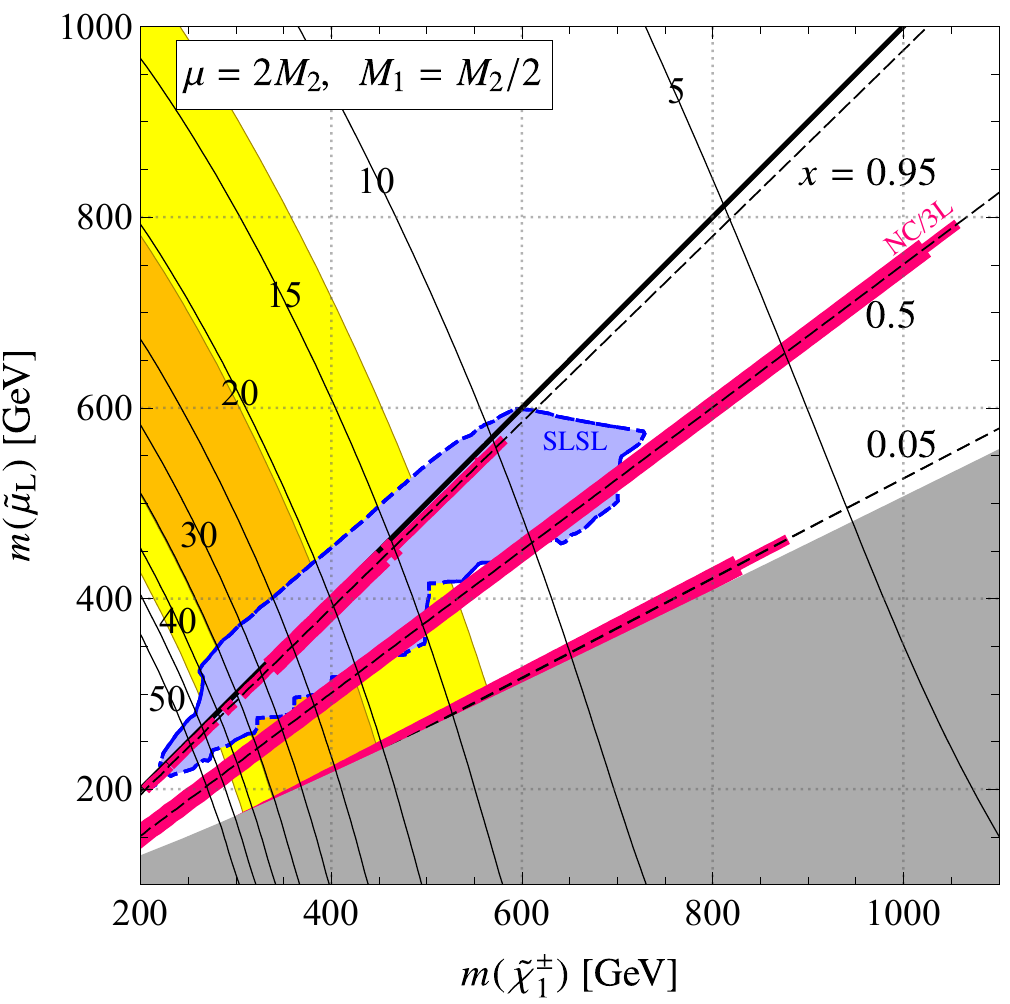}
\caption{$\mu = 2 M_2$, $M_1 = M_2/2$}\label{fig:2gut}
 \vspace{.2cm}
 \end{subfigure}
   \begin{subfigure}[b]{0.49\textwidth}
 \includegraphics[width=\textwidth]{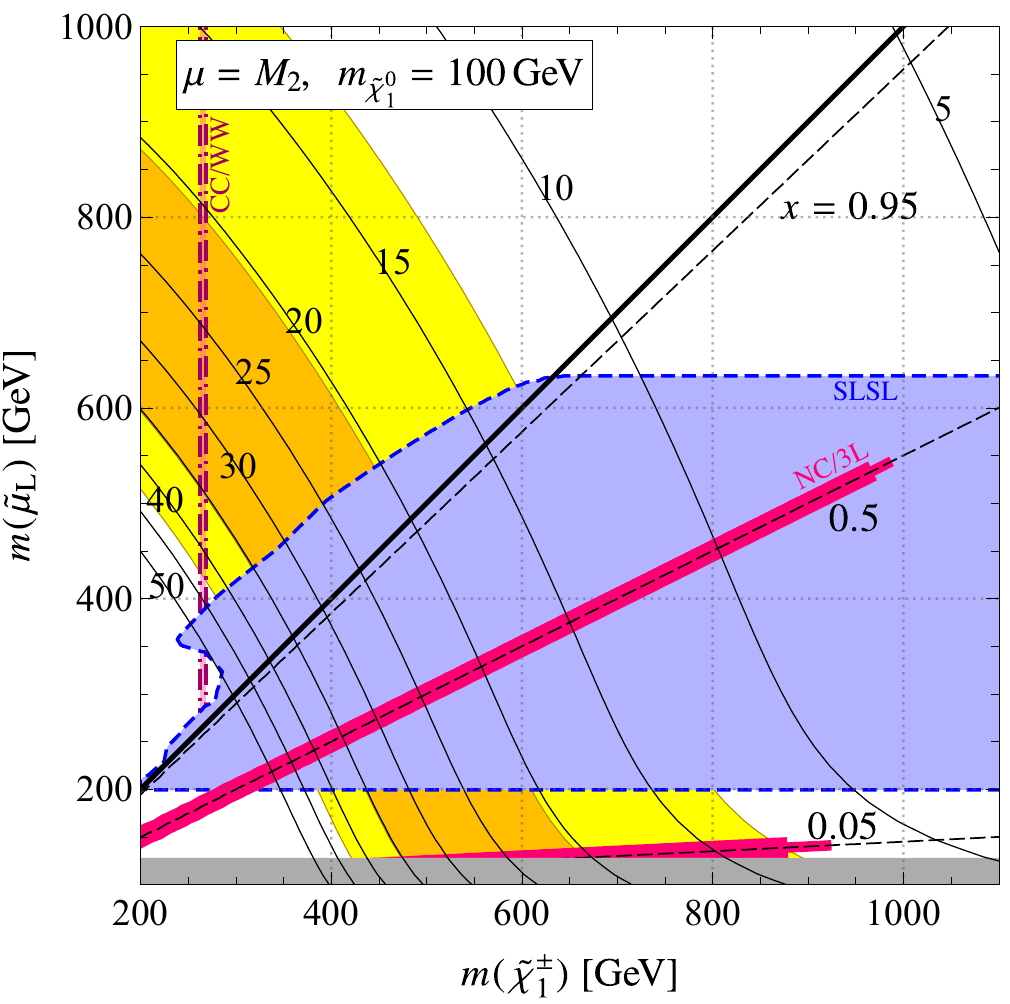}
\caption{$\mu = M_2$, $m_{\tilde{\chi}^0_1}=100$\,GeV}\label{fig:1m100}
 \end{subfigure}
   \begin{subfigure}[b]{0.49\textwidth}
 \includegraphics[width=\textwidth]{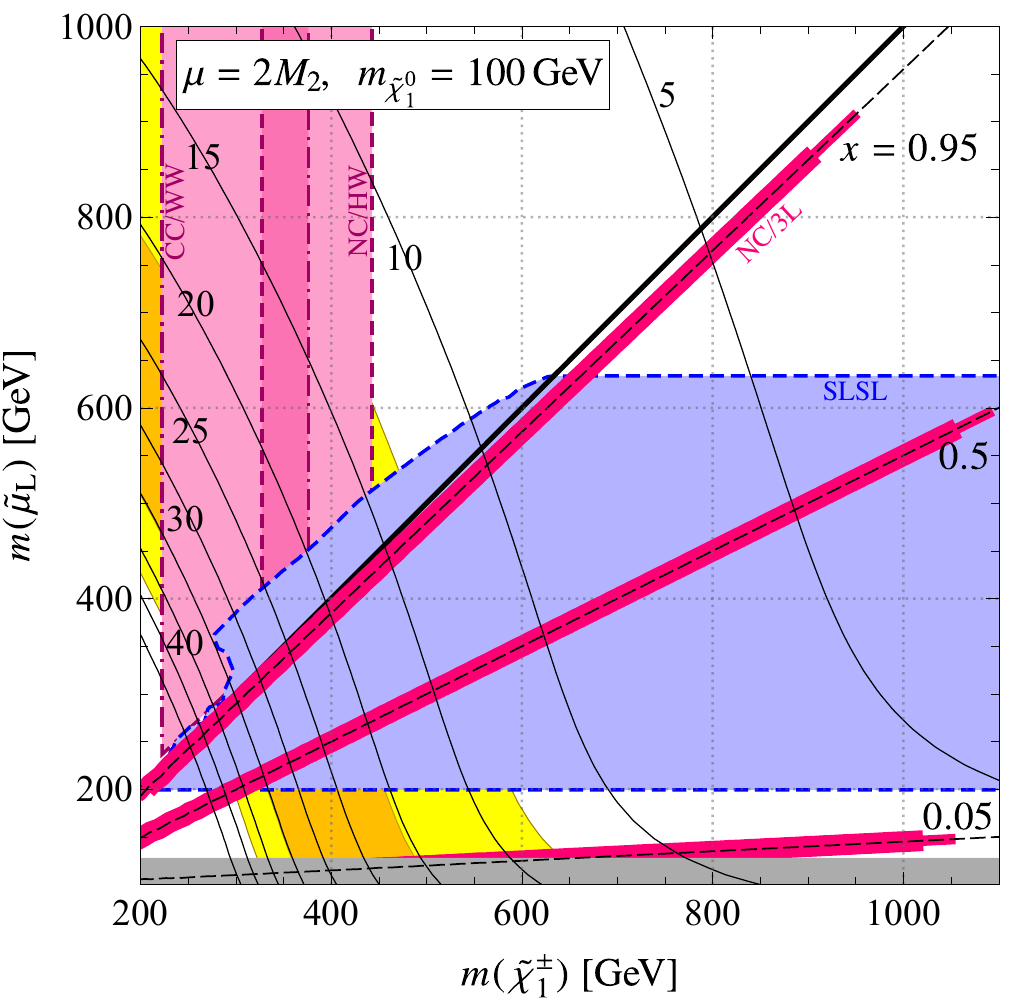}
\caption{$\mu = 2 M_2$, $m_{\tilde{\chi}^0_1}=100$\,GeV}\label{fig:2m100}
 \end{subfigure}
  \caption{\label{fig:summary}%
  \UPDATED{An updated version of this figure is available in Appendix~\ref{sec:update}.}
  LHC Run~2 bounds on the chargino-dominated SUSY scenario for the muon $g-2$ anomaly.
  Four parameter spaces with $\tan\beta=40$, Eq.~\eqref{eq:parameterspace}, are considered.
  The black contours show $\amu[SUSY]\times10^{10}$, but lines corresponding to $>50$ are omitted; $\amu[SUSY]=(27.8\pm7.4)\times10^{-10}$ is satisfied in the orange-filled (yellow-filled) regions at the $1\sigma$ ($2\sigma$) level.
  The thick black line corresponds to 
  $m_{\tilde{\mu}\w L} = m_{\charPM[1]} $.
  The gray-filled region, where the LSP is $\tilde\nu$, and the red-hatched region in (A), which corresponds to a compressed spectrum (see the text), are not studied.
  The LHC constraint from the CC/WW (NC/HW) analysis is shown by the red-filled regions with the dash-dotted (dashed) boundaries.
  The blue-filled regions are excluded by the SLSL analysis.
  The constraints from the NC/3L analysis are investigated on the model points with $x=0.05$, $0.5$, and $0.95$ (see Eq.~\eqref{eq:x}), where the exclusion ranges are shown by the magenta lines.
  }
\end{figure}

Our main results are summarized in Fig.~\ref{fig:summary}, where $\amu[SUSY]$ and the constraints from the LHC Run~2 are shown.\footnote{All the limits are given at the 95\% confidence level in this paper.}
The horizontal and vertical axes are the physical masses of the lighter chargino $\charPM[1]$ and the left-handed smuon $\tilde\mu\w L$, respectively.
The SUSY contribution to the muon $g-2$ is shown by the black solid contours in terms of $\amu[SUSY]\times10^{10}$; the results are shown up to $\amu[SUSY]=50\times10^{-10}$ and further contours are omitted for visibility.
In addition, the parameter spaces where $\amu[SUSY]$ solves the discrepancy $\Delta\amu=(27.8\pm7.4)\times10^{-10}$ at the $1\sigma$ ($2\sigma$) level are shown by the orange-filled (yellow-filled) regions.
The LHC constraints are shown by blue-filled regions, red-filled regions, and magenta lines. 
We discuss each of the constraints in the following section.

We do not address the relic abundance of the LSP, but concentrate on the parameter region where the lightest neutralino is the LSP.
The region with the sneutrino LSP is displayed in the plots by gray-filled region.

The red-hatched region in Fig.~\ref{fig:1gut} shows the parameter space in which all the two-body decays of $\neut[2]$ and/or $\charPM[1]$ are kinematically forbidden.
LHC constraints are strict in such cases and the region is expected to be excluded \cite{Endo:2013bba},
but definitive conclusion requires dedicated analysis with Monte Carlo simulation.
Hence, we do not discuss the region in this work.

\section{LHC bounds}
\label{sec:LHC}
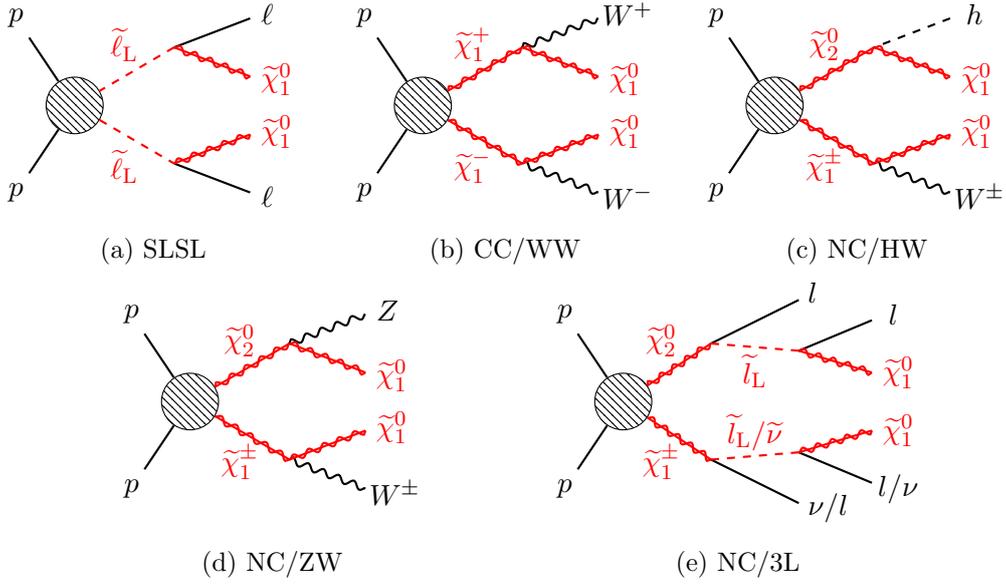
\begin{figure}[t]
  \centering
    \begin{subfigure}[b]{0.30\textwidth}
      \begin{minipage}{\linewidth}
    \centering
    \begin{tikzpicture}
      \begin{feynman}
        \vertex at ($(0em,-3.0em)$) (p1) {$p$};
        \vertex at ($(0em, 3.0em)$) (p2) {$p$};
        \vertex at ($(2em,   0em)$) [blob, draw=black, pattern color=black] (v){};
        \vertex at ($(v) + (3.4em, 2em)$) (d0);
        \vertex at ($(v) + (3.4em,-2em)$) (e0);
        \vertex at ($(v) + (6em,  3em)$) (d1);
        \vertex at ($(v) + (6em,  1em)$) (d2);
        \vertex at ($(v) + (6em, -1em)$) (e2);
        \vertex at ($(v) + (6em, -3em)$) (e1);
        \vertex at ($(v) + (1.7em, 2.1em)$)   (label1) {\color{red}$\tilde{\ell}\w L$};
        \vertex at ($(v) + (1.7em,-2.1em)$)   (label2) {\color{red}$\tilde{\ell}\w L$};
        \vertex at ($(d1) + (1.0em, 0.15em)$)   (label3) {$\ell$~~\,};
        \vertex at ($(d2) + (0.9em, -0.1em)$)  (label4) {\color{red}$\neut[1]$~};
        \vertex at ($(e2) + (0.9em, +0.1em)$)  (label5) {\color{red}$\neut[1]$~};
        \vertex at ($(e1) + (1.0em, -0.15em)$)  (label6) {$\ell$~~\,};
        \vertex at ($(v) + (0em, 3.6em)$)  (spacekeeper1) {};
        \vertex at ($(v) + (0em,-3.6em)$)  (spacekeeper2) {};
        \diagram*{
          (p1)--(v)--(p2),
          (d0)--[scalar,red] (v)--[scalar,red] (e0),
          (d1)--(d0)--[plain,boson,red](d2),
          (e1)--(e0)--[plain,boson,red](e2),
        };
      \end{feynman}
    \end{tikzpicture}
  \end{minipage}
  \caption{SLSL}\label{fig:SLSL}
 \end{subfigure}
      \begin{subfigure}[b]{0.30\textwidth}
        \begin{minipage}{\linewidth}
    \centering
    \begin{tikzpicture}
      \begin{feynman}
        \vertex at ($(0em,-3.0em)$) (p1) {$p$};
        \vertex at ($(0em, 3.0em)$) (p2) {$p$};
        \vertex at ($(2em,   0em)$) [blob, draw=black, pattern color=black] (v){};
        \vertex at ($(v) + (3.4em, 2em)$) (d0);
        \vertex at ($(v) + (3.4em,-2em)$) (e0);
        \vertex at ($(v) + (6em,  3em)$) (d1);
        \vertex at ($(v) + (6em,  1em)$) (d2);
        \vertex at ($(v) + (6em, -1em)$) (e2);
        \vertex at ($(v) + (6em, -3em)$) (e1);
        \vertex at ($(v) + (1.7em, 2.1em)$)   (label1) {\color{red}$\tilde{\chi}^+_1$};
        \vertex at ($(v) + (1.7em,-2.1em)$)   (label2) {\color{red}$\tilde{\chi}^-_1$};
        \vertex at ($(d1) + (1.0em, 0.15em)$)   (label3) {$W^+$};
        \vertex at ($(d2) + (0.9em, -0.1em)$)  (label4) {\color{red}$\neut[1]$~};
        \vertex at ($(e2) + (0.9em, +0.1em)$)  (label5) {\color{red}$\neut[1]$~};
        \vertex at ($(e1) + (1.0em, -0.15em)$)  (label6) {$W^-$};
        \vertex at ($(v) + (0em, 3.6em)$)  (spacekeeper1) {};
        \vertex at ($(v) + (0em,-3.6em)$)  (spacekeeper2) {};
        \diagram*{
          (p1)--(v)--(p2),
          (d0)--[plain, boson, red] (v)--[plain, boson, red] (e0),
          (d1)--[boson](d0)--[plain,boson,red](d2),
          (e1)--[boson](e0)--[plain,boson,red](e2),
        };
      \end{feynman}
    \end{tikzpicture}
  \end{minipage}
    \caption{CC/WW}\label{fig:CC/WW}
 \end{subfigure}
      \begin{subfigure}[b]{0.30\textwidth}
     \begin{minipage}{\linewidth}
    \centering
    \begin{tikzpicture}
      \begin{feynman}
        \vertex at ($(0em,-3.0em)$) (p1) {$p$};
        \vertex at ($(0em, 3.0em)$) (p2) {$p$};
        \vertex at ($(2em,   0em)$) [blob, draw=black, pattern color=black] (v){};
        \vertex at ($(v) + (3.4em, 2em)$) (d0);
        \vertex at ($(v) + (3.4em,-2em)$) (e0);
        \vertex at ($(v) + (6em,  3em)$) (d1);
        \vertex at ($(v) + (6em,  1em)$) (d2);
        \vertex at ($(v) + (6em, -1em)$) (e2);
        \vertex at ($(v) + (6em, -3em)$) (e1);
        \vertex at ($(v) + (1.7em, 2.1em)$)   (label1) {\color{red}$\tilde{\chi}^0_2$};
        \vertex at ($(v) + (1.7em,-2.1em)$)   (label2) {\color{red}$\tilde{\chi}^\pm_1$};
        \vertex at ($(d1) + (1.0em, 0.15em)$)   (label3) {$h$~~};
        \vertex at ($(d2) + (0.9em, -0.1em)$)  (label4) {\color{red}$\neut[1]$~};
        \vertex at ($(e2) + (0.9em, +0.1em)$)  (label5) {\color{red}$\neut[1]$~};
        \vertex at ($(e1) + (1.0em, -0.15em)$)  (label6) {$W^\pm$};
        \vertex at ($(v) + (0em, 3.6em)$)  (spacekeeper1) {};
        \vertex at ($(v) + (0em,-3.6em)$)  (spacekeeper2) {};
        \diagram*{
          (p1)--(v)--(p2),
          (d0)--[plain, boson, red] (v)--[plain, boson, red] (e0),
          (d1)--[scalar](d0)--[plain,boson,red](d2),
          (e1)--[boson](e0)--[plain,boson,red](e2),
        };
      \end{feynman}
    \end{tikzpicture}
  \end{minipage}
    \caption{NC/HW}\label{fig:NC/HW}
 \end{subfigure}
      \begin{subfigure}[b]{0.4\textwidth}
       \begin{minipage}{\linewidth}
    \centering
    \begin{tikzpicture}
      \begin{feynman}
        \vertex at ($(0em,-3.0em)$) (p1) {$p$};
        \vertex at ($(0em, 3.0em)$) (p2) {$p$};
        \vertex at ($(2em,   0em)$) [blob, draw=black, pattern color=black] (v){};
        \vertex at ($(v) + (3.4em, 2em)$) (d0);
        \vertex at ($(v) + (3.4em,-2em)$) (e0);
        \vertex at ($(v) + (6em,  3em)$) (d1);
        \vertex at ($(v) + (6em,  1em)$) (d2);
        \vertex at ($(v) + (6em, -1em)$) (e2);
        \vertex at ($(v) + (6em, -3em)$) (e1);
        \vertex at ($(v) + (1.7em, 2.1em)$)   (label1) {\color{red}$\tilde{\chi}^0_2$};
        \vertex at ($(v) + (1.7em,-2.1em)$)   (label2) {\color{red}$\tilde{\chi}^\pm_1$};
        \vertex at ($(d1) + (1.0em, 0.15em)$)   (label3) {$Z$~\,};
        \vertex at ($(d2) + (0.9em, -0.1em)$)  (label4) {\color{red}$\neut[1]$~};
        \vertex at ($(e2) + (0.9em, +0.1em)$)  (label5) {\color{red}$\neut[1]$~};
        \vertex at ($(e1) + (1.0em, -0.15em)$)  (label6) {$W^\pm$};
        \vertex at ($(v) + (0em, 4.2em)$)  (spacekeeper1) {};
        \vertex at ($(v) + (0em,-4.2em)$)  (spacekeeper2) {};
        \diagram*{
          (p1)--(v)--(p2),
          (d0)--[plain, boson, red] (v)--[plain, boson, red] (e0),
          (d1)--[boson](d0)--[plain,boson,red](d2),
          (e1)--[boson](e0)--[plain,boson,red](e2),
        };
      \end{feynman}
    \end{tikzpicture}
  \end{minipage}
    \caption{NC/ZW}\label{fig:NC/ZW}
 \end{subfigure}
      \begin{subfigure}[b]{0.4\textwidth}
  \begin{minipage}{\linewidth}
    \centering
  \begin{tikzpicture}
    \begin{feynman}
      \vertex at ($(0em,-3.0em)$) (p1) {$p$};
      \vertex at ($(0em, 3.0em)$) (p2) {$p$};
      \vertex at ($(2em,   0em)$) [blob, draw=black, pattern color=black] (v){};
      \vertex at ($(v) + (3em, 2em)$) (d0);
      \vertex at ($(v) + (3em,-2em)$) (e0);
      \vertex at ($(v) + (6em,  3.5em)$) (d1);
      \vertex at ($(v) + (6em,  1.75em)$) (d2);
      \vertex at ($(v) + (6em, -1.75em)$) (e2);
      \vertex at ($(v) + (6em, -3.5em)$) (e1);
      \vertex at ($(v) + (8.5em,  2.8em)$) (d2a);
      \vertex at ($(v) + (8.5em,  1em)$) (d2b);
      \vertex at ($(v) + (8.5em, -1em)$) (e2b);
      \vertex at ($(v) + (8.5em, -2.8em)$) (e2a);

      \vertex at ($(v) + (1.3em, 2.1em)$)   (label1) {\color{red}$\tilde{\chi}^0_2$};
      \vertex at ($(v) + (1.3em,-2.1em)$)   (label2) {\color{red}$\tilde{\chi}^\pm_1$};
      \vertex at ($(d0) + (1.5em, -0.95em)$) (label3) {\color{red}$\tilde{l}\w L$};
      \vertex at ($(e0) + (1.5em,  0.95em)$) (label4) {\color{red}${\tilde{l}\w L}/{\tilde{\nu}}$};
      \vertex at ($(d1) + (1em, 0.15em)$)   (label5) {$l\phantom{/\nu}$};
      \vertex at ($(e1) + (1em, -0.15em)$)  (label6) {$\nu/l$};
      \vertex at ($(v) + (9.4em,  3em)$) (label7){$l$~~};
      \vertex at ($(v) + (9.4em,  1em)$) (label8){\color{red}$\neut[1]$~};
      \vertex at ($(v) + (9.4em, -1em)$) (label9){\color{red}$\neut[1]$~};
      \vertex at ($(v) + (9.4em, -3em)$) (label10){$l/\nu$};
      \vertex at ($(v) + (0em, 4.2em)$)  (spacekeeper1) {};
      \vertex at ($(v) + (0em,-4.2em)$)  (spacekeeper2) {};
      \diagram*{
            (p1)--(v)--(p2),
             (d0)-- [plain, boson, red] (v) -- [plain, boson, red] (e0),
             (d1)--(d0)--[scalar,red](d2),
             (e1)--(e0)--[scalar,red](e2),
             (d2a)--(d2)--[plain,boson,red](d2b),
             (e2a)--(e2)--[plain,boson,red](e2b),
      };
    \end{feynman}
  \end{tikzpicture}
  \end{minipage}
    \caption{NC/3L}\label{fig:NC/3L}
 \end{subfigure}
  \caption{
  Feynman diagrams of the processes we investigate.
  Here, $\ell = e,\mu$ and $l = e,\mu,\tau$.
}
  \label{fig:diagram}
\end{figure}
In the present setup, the neutralinos, charginos, and sleptons are the targets at the LHC.
We consider the following channels which have been studied by the ATLAS and CMS collaborations: 
\begin{align}
\text{SLSL:}\quad& pp\to \tilde \ell\w L \tilde \ell\w L^{\ast} \to 
(\ell \neut[1])(\bar \ell \neut[1])\,,
\\
\text{CC/WW:}\quad & pp\to \charP[1]\charM[1]\to
(W^{+}\tilde{\chi}^0_1)(W^{-}\tilde{\chi}^0_1)\,,\label{eq:CC/WW}\\
\text{NC/HW:}\quad & pp\to \neut[2]\charPM[1]\to(h\neut[1])(W^\pm\neut[1])\,,\label{eq:NC/HW}\\
\text{NC/ZW:}\quad & pp\to \neut[2]\charPM[1]\to(Z\neut[1])(W^\pm\neut[1])\,,\label{eq:NC/ZW}\\
\text{NC/3L:}\quad& pp\to \neut[2]\charPM[1]\to 
\begin{cases}
 (l \tilde l\w L)(\nu \tilde{l}\w L)
 \to
 (ll \neut[1])(\nu l \neut[1])\,,
 \\
  (l \tilde{l}\w L) (l \tilde{\nu})
 \to
 (l l \neut[1])(l \nu \neut[1])\,.
\end{cases}
\label{eq:NC/3L}
\end{align}
Here and hereafter, $\ell=e,\mu$, $\tilde{\ell}\w L=\tilde{e}\w L,\tilde{\mu}\w L$, $l=e,\mu,\tau$, $\tilde{l}\w L=\tilde{e}\w L,\tilde{\mu}\w L,\tilde{\tau}\w L$, and 
$\tilde{\nu}=\tilde{\nu}_{e,\textrm{L}},\tilde{\nu}_{\mu,\textrm{L}},\tilde{\nu}_{\tau,\textrm{L}}$.
The corresponding diagrams are exhibited in Fig.~\ref{fig:diagram}.
Although more channels have been studied by the experiment collaborations, we found that they are less sensitive under our setup.
See also Appendices~\ref{app:Auxiliary} and \ref{app:LHC_analysis_list} for details of the analysis of the LHC constraints and the summary of the LHC Run~2 results investigated in our study, respectively.

Since the electroweakinos are mixtures of the bino, wino, and higgsino gauge eigenstates,
the production cross sections and the decay patterns are determined by the mixing composition and thus dependent on the parameters in Eq.~\eqref{eq:parameters}.
This is contrasted to typical setups in analyses of the ATLAS and CMS collaborations. 
LHC constraints are usually reported on simplified models, \eg, some of the branching ratios are fixed to be unity. %
Therefore,  to obtain LHC constraints on the SUSY models motivated by the muon $g-2$ anomaly, we need to interpret the LHC constraints in terms of the model parameters of our interest (cf.~Ref.~\cite{Endo:2013bba}).

The ATLAS and CMS collaborations usually report upper limits (UL) on production cross sections, $\sigma\UL$, in a simplified scenario.
When we focus on a specific signal region (SR), it is related to the upper limit on the number of events in this SR, $N\UL$, by
\begin{equation}
 \frac{N\UL}{\int\! dt \mathcal L} = (\mathcal A\times\mathcal E)|\w{original}\cdot\sigma\ULorig\,,
\label{eq:NUL}
\end{equation}
where $\int\! dt \mathcal L$ is the integrated luminosity, $\mathcal A$ the acceptance, and $\mathcal E$ the efficiency.
A label ``original'' is introduced to clarify that the values are for the original simplified scenario by the ATLAS and CMS collaborations.
The left-hand side is independent of the processes or models. The dependence is contained in the right-hand side.
Accordingly, the constraints can be applied to any model $X$ by calculating the acceptance and efficiency.
The upper limit on the production cross section under the model, $\sigma\ULX$, is derived as
\begin{equation}
 \sigma\ULX =
 \frac{(\mathcal A\times\mathcal E)|\w{original}}{(\mathcal A\times\mathcal E)|_{X}}\cdot\sigma\ULorig\,.
\label{eq:ULtranslation}
\end{equation}

Although the acceptance and efficiency, $(\mathcal A\times\mathcal E)|_{X}$, may be estimated by Monte Carlo simulation with detector simulation, it is not straightforward to recast the experimental results in generic models  because recent LHC analyses become more and more involved.
The results obtained by Monte Carlo simulation are subject to extra uncertainties.
In Ref.~\cite{Endo:2013bba}, we assigned an extra 30\% uncertainty, though the number was without rigorous justification.
Therefore, even if we emulate the LHC searches by ourselves, we are not able to evaluate the credibility of the results in a quantitative manner. The results should be regarded only as a qualitative guideline.
In this work, we refrain from emulating the LHC searches by Monte Carlo simulation in order to avoid such extra uncertainties.
Instead, we reinterpret the reported LHC constraints in our model points directly. 
If the kinematics of the process in our model point imitates the original one, \ie, when the mass spectrum of the SUSY particles relevant for the SR is the same as that in the LHC analysis, we recast the experimental results by approximating the ratio as
\begin{equation}
 \frac{(\mathcal A\times\mathcal E)|\w{original}}{(\mathcal A\times\mathcal E)|_{X}}
\approx
 \frac{B\w{original}}{B_X}\,,
\label{eq:ULapprox}
\end{equation}
where the right-hand side is estimated from branching ratios of the final-state particles.
See the following subsections for explicit cases.
Note that we do not combine the constraints from multiple SRs but study them separately;
the interpreted constraints will be conservative, \ie, weaker than the ones we could obtain with full emulation, but free from extra uncertainty and justifiable.
Then, the model is excluded if the theoretical production cross section $\sigma_X$ satisfies
\begin{equation}
 \sigma_X > \sigma\ULX=\frac{B\w{original}}{B_X} \,\sigma\ULorig \,.
 \label{eq:bound}
\end{equation}
In the following, we explain the LHC bound for each channel.

\subsection{Slepton pair-production (SLSL)}\label{sec:SLSL}

Model points with light sleptons are constrained by searching for pair-productions of the sleptons (\figref{SLSL}):\footnote{%
The process $pp\to\tilde\tau\w L\tilde\tau\w L^*$ is searched for by the two hadronic taus plus $\mET$ signature, but no constraint is obtained for $m_{\neut[1]} > 100\,$GeV \cite{Aad:2019byo}.
Also, its contribution to the $2\ell+\mET$ signature is negligible because of the smaller branching ratio.
}
\begin{equation}
 p p\to \tilde{\ell}\w L \tilde{\ell}^{\ast}\w L \to \ell \tilde{\chi}^0_1  \bar\ell \tilde{\chi}^0_1\,.
 \label{eq:processSLSL}
\end{equation}
Currently, the ATLAS collaboration provides the most severe constraint by studying the two leptons plus missing transverse energy signature ($2 \ell + \mET$) at $\sqrt{s}=13\TeV$ and $\int\! dt \mathcal{L}=139\ifb$ \cite{Aad:2019vnb}.
In the ATLAS analysis, it is assumed that $\tilde{e}\w L, \tilde{e}\w R, \tilde{\mu}\w L$, and $\tilde{\mu}\w R$ are mass-degenerate and that each slepton decays into a lepton and the lightest neutralino with a 100\% branching ratio. In our setup, only the left-handed sleptons are light, and $\mathrm{Br}(\tilde{\ell} \to \ell \tilde{\chi}^0_1) \neq 1$ because
$\tilde{\ell} \to \nu_\ell  \tilde{\chi}^-_1$ can be open.
Hence,
\begin{equation}
  \frac{B_X}{B\w{original}} = \Br(\tilde\ell\w L\to\ell\neut[1])^2
\end{equation}
is used to compare the experimental bound with the theoretical production cross section as stated in Eq.~\eqref{eq:ULapprox}.\footnote{For the process in Eq.~\eqref{eq:processSLSL}, although the full acceptance and efficiency should also depend on kinematics of the final states, SR cuts, and details of the detectors, all of these factors are the same as in the ATLAS analysis if masses of the slepton and the lightest neutralino are the same as their setup.
Consequently, all the factors except for the branching ratio cancel out in the ratio of $(\mathcal A\times\mathcal E)$ in Eq.~\eqref{eq:ULtranslation}.}
The experimental bound, $\sigma\ULorig$, is provided in  Ref.~\cite{10.17182/hepdata.89413.v1/t47} and the cross section, $\sigma_X$, is available with the NLO-NLL accuracy~\cite{LHCSUSYCSWG,Beenakker:1999xh,Bozzi:2007qr,Fuks:2013vua,Fuks:2013lya,Fiaschi:2018xdm}.

In Fig.~\ref{fig:summary}, the LHC constraints from the SLSL channel are shown by the blue-filled regions.
For $m_{\tilde{\mu}\w L} \gtrsim m_{\charPM[1]}$, the decay channel $\tilde{\ell} \to \nu_\ell  \tilde{\chi}^-_1$ is open, and thus, $B_X/B\w{original}\ll1$, \ie, the events are less accepted than those in the ATLAS simplified model, so that weaker constraints are obtained.
Note that further regions could be excluded by taking account of the slepton cascade decays, \eg,
$\tilde{\ell} \to \nu_\ell  \tilde{\chi}^-_1 \to \nu_\ell W^- \tilde{\chi}^0_1 \to \nu_\ell \ell' \nu_{\ell'} \tilde{\chi}^0_1 $. However, we do not discuss them because definitive conclusion requires dedicated study with Monte Carlo simulation.

For $m_{\tilde{\mu}\w L} < m_{\charPM[1]}$, $B_X/B\w{original}\simeq 1$ and thus the masses of $\tilde\ell\w{L}$ and $\neut[1]$ determine the constraint.
In the cases of $M_1 = M_2/2$, the constraint is loosened for heavier $\charPM[1]$ because $m_{\tilde\ell}$ gets closer to $m_{\neut[1]}$. In the cases of $M_1 = 100$\,GeV, \figsref{1m100}{2m100}, the regions with $200\GeV< m_{\tilde\ell\w L}<634\GeV$ are excluded.\footnote{%
Although we show limits with $1\GeV$ precision here and hereafter, the last digits should not be considered seriously because we interpolate the upper bounds provided by the ATLAS and CMS collaborations.
}
The regions of $m_{\tilde\ell\w L}<200\GeV$ and $>634\GeV$ are not excluded by this channel because of the mass degeneracy and the smaller production cross section, respectively.

\subsection{Chargino pair-production (CC/WW)}\label{sec:CC/WW}
The next channel is the chargino pair-production, $pp\to\charP[i]\charM[j]$.
Let us first focus on the production of the lightest pair, $\charP[1]\charM[1]$.
If $\charPM[1]$ is heavier than $\tilde l\w L$, the charginos decay via sleptons,
\begin{equation}\label{eq:CC2L}
 pp\to\charP[1]\charM[1]\to
 (\nu\tilde l^*\w L~\text{or}~\bar l \tilde{\nu})
 (\bar\nu \tilde{l}\w L~\text{or}~l \tilde{\nu}^*)
 \to
 (\nu \bar l \neut[1])(\bar\nu l \neut[1])\,,
\end{equation}
which provides the $2\ell+\mET$ signature (cf.~Ref.~\cite{Aad:2019vnb}).
However, such models also include the NC/3L process, Eq.~\eqref{eq:NC/3L}, which we discuss in Sec.~\ref{sec:NC/3L}.
The constraints from the NC/3L process generally give stronger constraints because it provides more characteristic signature such as the $3\ell+\mET$ or same-sign $2\ell$ plus $\mET$ signatures and also because it has a larger cross section (see \figref{pure-ino-xs} in Appendix~\ref{sec:AppA1}).
We have checked that the constraints from Eq.~\eqref{eq:CC2L} are weaker than those from the NC/3L in our model points and we do not discuss those further.

If $\charPM[1]$ is lighter than $\tilde l\w L$, two $W$ bosons are generated from charginos (\figref{CC/WW}),
\begin{equation}
pp\to \charP[1]\charM[1]\to (W^{+}\tilde{\chi}^0_1)(W^{-}\tilde{\chi}^0_1)\,.
\end{equation}
With $\sqrt{s}=13\TeV$ and $\int\! dt \mathcal{L}=139\ifb$, the most severe constraint is given by the ATLAS collaboration, where the $2 \ell + \mET$ signature is studied \cite{Aad:2019vnb}.
In the ATLAS analysis, $\Br(\tilde{\chi}^{\pm}_1 \to W^{\pm} \tilde{\chi}^0_1) = 1$ is assumed.
In our setup, we obtain upper bounds on $\sigma(pp\to\charP[1]\charM[1])$ by using 
\begin{equation}
 \frac{B_X}{B\w{original}}=\Br(\tilde{\chi}^{\pm}_1 \to W^{\pm} \tilde{\chi}^0_1)^2\,,
 \label{eq:CCWW-B}
\end{equation}
which is approximately unity if $\charPM[1]$ is lighter than $\tilde l\w L$.
The ATLAS upper bounds are available in Ref.~\cite{10.17182/hepdata.89413.v1/t45}.
For the theoretical cross section, we calculate the chargino pair-production cross section at the tree level by \package[2.6.5]{MadGraph5\_aMC@NLO} \cite{Alwall:2011uj,Alwall:2014hca} with squarks decoupled. Then, the NLO-NLL cross section is estimated by multiplying the $K$-factor calculated from the NLO-NLL cross section 
\cite{LHCSUSYCSWG,Beenakker:1999xh,Debove:2010kf,Fuks:2012qx,Fuks:2013vua}. 
As a result, the cross section is approximated as
\begin{equation}
 \sigma_X \approx \frac{\sigma\w{tree}(pp\to\charP[1]\charM[1])}{\sigma\w{tree}(pp\to\tilde W^+\tilde W^-)}{\sigma_{\text{NLO-NLL}}(pp\to\tilde W^+\tilde W^-)}\,.
\label{eq:CCWWxsapprox}
\end{equation}

From Eqs.~\eqref{eq:CCWW-B} and \eqref{eq:CCWWxsapprox}, the ATLAS bound is reinterpreted for our models.
\Figref{c1c1summary} shows the results for the model points (C) and (D), while no constraints are obtained for (A) and (B).
The black solid lines show the theoretical cross section $\sigma_X$; its tree-level value, $\sigma\w{tree}(pp\to\charP[1]\charM[1])$, is also displayed by the gray dashed lines (see Eq.~\eqref{eq:CCWWxsapprox}).
In addition, the two black dotted lines show the theoretical cross sections in the pure-wino limit (the upper line) and in the pure-higgsino limit (the lower line).
The red lines are the upper bounds $\sigma\ULX$, which are applicable for model points with $m_{\charPM[1]}<m_{\tilde l\w L}$.
Consequently, the exclusion regions are obtained in \figsref{1m100}{2m100}, which are $273\GeV < m_{\charPM[1]} <279\,$GeV and $223\GeV < m_{\charPM[1]} <376\,$GeV, respectively, with $m_{\charPM[1]}<m_{\tilde l\w L}$.

\begin{figure}[t]
 \centering
 \begin{subfigure}[b]{0.48\textwidth}
  \includegraphics[width=\textwidth]{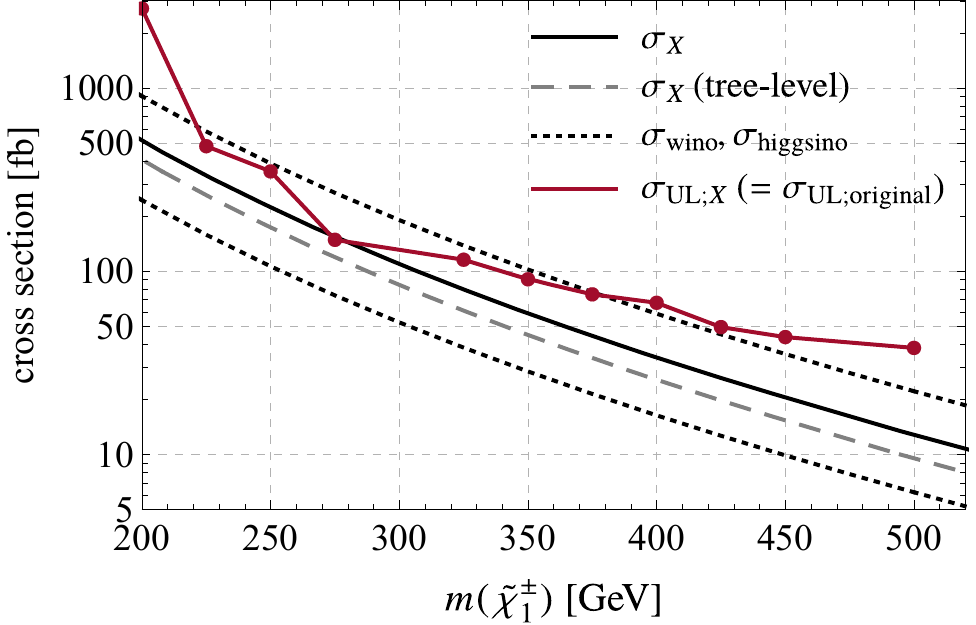}
  \caption{$\mu = M_2$, $m_{\tilde{\chi}^0_1}=100$\,GeV}
 \end{subfigure}\hspace{0.02\textwidth}
 \begin{subfigure}[b]{0.48\textwidth}
  \includegraphics[width=\textwidth]{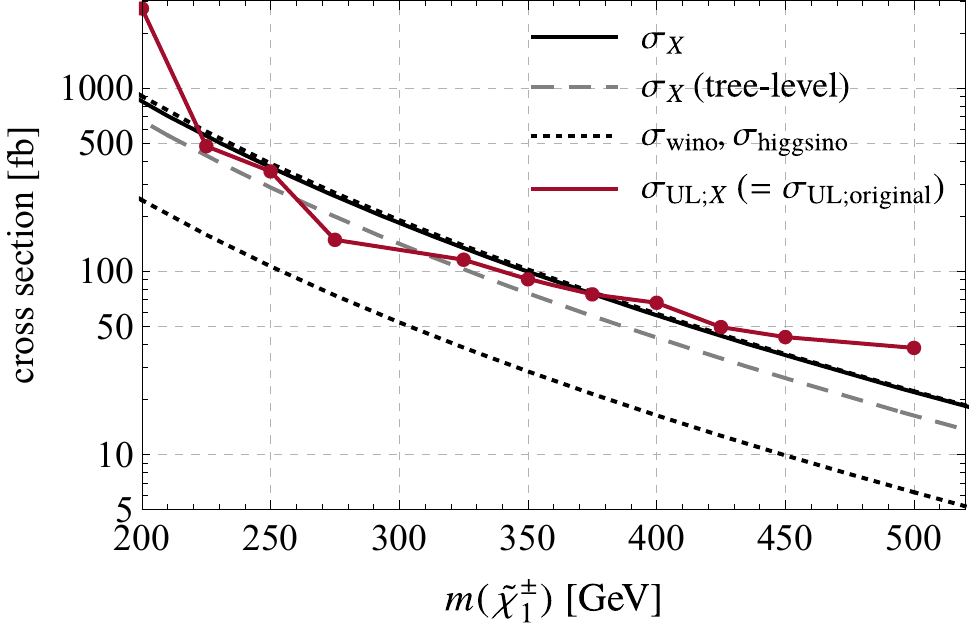}
  \caption{$\mu = 2M_2$, $m_{\tilde{\chi}^0_1}=100$\,GeV}
 \end{subfigure}
\caption{
  Summary plot of the CC/WW analysis.
  The theoretical cross section, $\sigma_X(pp\to\charP[1]\charM[1])$, is shown by the black solid line together with its tree-level values shown by gray dashed lines.
  The red solid line shows the experimental upper limit, $\sigma\ULX$, applicable if $m_{\tilde l\w L}>m_{\charPM[1]}$; it is obtained from the ATLAS result of the $2\ell+\mET$ signature~\cite{Aad:2019vvf,10.17182/hepdata.90607.v1/t17}.
  In addition, the chargino pair-production cross section in the pure-wino (pure-higgsino) limit is shown by the upper (lower) black dotted line.}
\label{fig:c1c1summary}
\end{figure}

The bound in \figref{1m100} is weaker than in \figref{2m100} because a larger higgsino component in $\charPM[1]$ suppresses the production cross section, $\sigma(p p\to \tilde{\chi}^+_1 \tilde{\chi}^{-}_1)$.
In fact, $\charPM[1]$ is wino-like in the model (D) due to $\mu=2M_2$, and thus $\sigma_X$ in the right panel of \figref{c1c1summary} becomes close to the pure-wino value.
On the other hand, $\charPM[1]$ in the model (C) is a wino-higgsino mixture, and $\sigma_X$ is amid the pure-wino and pure-higgsino values.
Note that, if $\mu=2M_2$ and $m_{\charPM[1]}<m_{\tilde l\w L}$, our setup becomes almost equivalent to the ATLAS simplified model.

For both of \figsref{1m100}{2m100}, the regions with $m_{\charPM[1]}\sim200\GeV$ are not yet constrained.
This is because the mass separation between $\charPM[1]$ and $\neut[1]$ is small.
There are no excluded regions in \figsref{1gut}{2gut} for the same reason.
Additional information is provided in Appendix~\ref{app:EWK}.

Let us briefly mention the process other than $pp\to\charP[1]\charM[1]$, \ie, contributions from the heavier chargino $\charPM[2]$.
As shown in Appendix~\ref{app:EWK}, they exhibit cascade decays, \eg, $\charP[2]\to Z\charP[1]\to Z W\neut[1]$.
Although the cascade may provide characteristic collider signature with small SM backgrounds, such analyses have not been performed yet. As they require Monte Carlo simulation, we do not discuss them in this work.

\subsection%
[Electroweakino pair-production in final state with hW (NC/HW)]%
{Electroweakino pair-production in final state with $hW$ (NC/HW)}\label{sec:NC/HW}

The neutralino-chargino pair-production, $pp\to\neut[i]\charPM[j]$, which exhibits various decay signatures, has been studied in various analyses. Important bounds are obtained because the cross section is large (see Fig.~\ref{fig:pure-ino-xs} in Appendix~\ref{sec:AppA1}).
For our model points, three signatures are expected to give decisive constraints, which we will discuss in this and the following subsections.
The first channel is (\figref{NC/HW})
\begin{equation}
p p \to \neut[2]\charPM[1] \to 
h \tilde{\chi}^0_1 W^{\pm} \tilde{\chi}^0_1\,.
\end{equation}
The LHC Run~2 results are reported in Refs.~\cite{Aad:2019vvf,Sirunyan:2017lae,Sirunyan:2018ubx}. The leading constraint is given by the ATLAS collaboration with $\int\! dt \mathcal{L}=139\ifb$, where the $1\ell b \bar{b} + \mET$ signature is studied with $\Br(\tilde{\chi}^{0}_2 \to h \tilde{\chi}^0_1)
=  \Br(\tilde{\chi}^{\pm}_1 \to W^{\pm} \tilde{\chi}^0_1)=1$ \cite{Aad:2019vvf}.
The result can be interpreted with
\begin{equation}
  \frac{B_X}{B\w{original}}=
  \Br(\tilde{\chi}^{0}_2 \to h \tilde{\chi}^0_1)
  \Br(\tilde{\chi}^{\pm}_1 \to W^{\pm} \tilde{\chi}^0_1)\,.
  \label{eq:NCHW-B}
\end{equation}
This is sizable only if the electroweakinos are lighter than sleptons; otherwise they mostly decay into sleptons and thus the NC/HW channel becomes irrelevant.

The ATLAS upper bound, $\sigma\UL(p p\to \tilde{\chi}^{0}_2 \tilde{\chi}^{\pm}_1)$, is available in Ref.~\cite{10.17182/hepdata.90607.v1/t17}.
As we show in Appendix~\ref{app:xs}, the tree-level cross section is approximately given by
\begin{equation}
\frac{\sigma\w{tree}(pp\to\neut[2]\charPM[1])}{\sigma\w{tree}(pp\to\tilde W^0\tilde W^\pm)}\approx
R^{\mathrm{NC}}_{21}\,,
\label{eq:Xsecchi21}
\end{equation}
where $R^{\mathrm{NC}}_{21}$ is defined in Eq.~\eqref{eq:RNCdef} as an analytic function of the electroweakino mixings.
We thus approximate the theoretical cross section as
\begin{equation}
 \sigma_X(pp\to\neut[2]\charPM[1]) \approx R^{\mathrm{NC}}_{21}\times
{\sigma_{\text{NLO-NLL}}(pp\to\tilde W^0\tilde W^\pm)}\,,
\label{eq:N2C1production}
\end{equation}
where the pure-wino cross section, $\sigma_{\text{NLO-NLL}}(pp\to\tilde W^0\tilde W^\pm)$, is provided in Ref.~\cite{LHCSUSYCSWG} (cf.~Refs.~\cite{Beenakker:1999xh,Debove:2010kf,Fuks:2012qx,Fuks:2013vua,Fiaschi:2018hgm}).

\begin{figure}[t]
 \renewcommand\thesubfigure{\Alph{subfigure}}
 \centering
 \begin{subfigure}[t]{0.48\textwidth}
  \includegraphics[width=\textwidth]{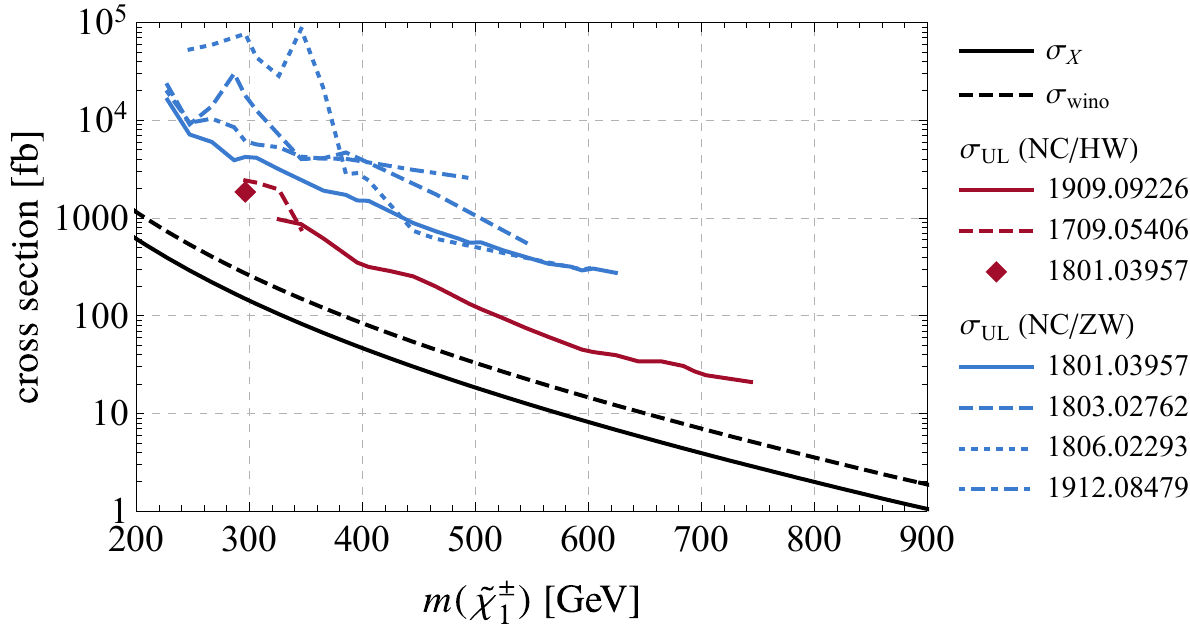}
  \caption{$\mu = M_2$, $M_1=M_2/2$}\label{fig:n2c1summaryA}
 \end{subfigure}\hspace{0.02\textwidth}
 \begin{subfigure}[t]{0.48\textwidth}
  \includegraphics[width=\textwidth]{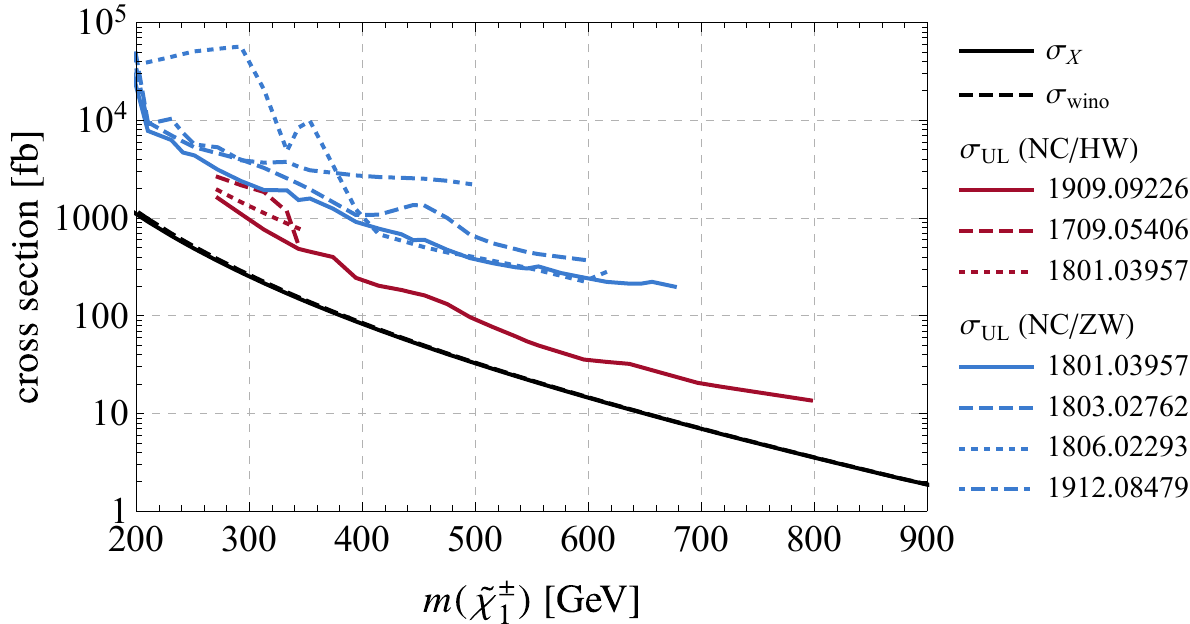}
  \caption{$\mu = 2M_2$, $M_1=M_2/2$}\label{fig:n2c1summaryB}
 \end{subfigure}
\par\vspace{1em}
 \begin{subfigure}[t]{0.48\textwidth}
  \includegraphics[width=\textwidth]{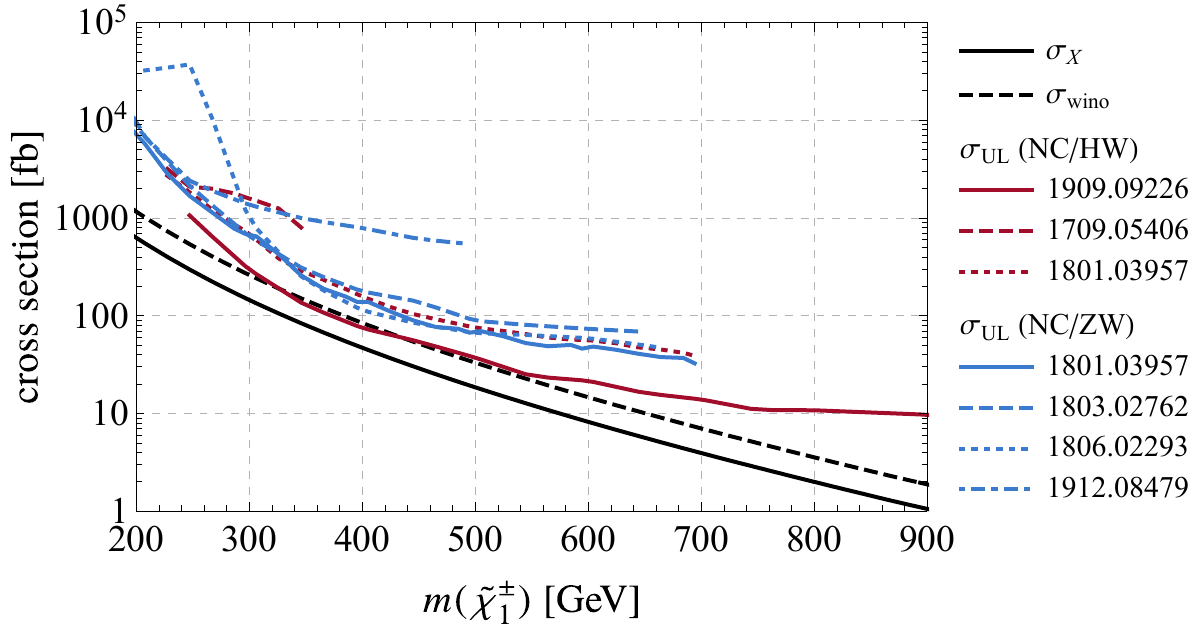}
  \caption{$\mu = M_2$, $m_{\tilde{\chi}^0_1}=100$\,GeV}\label{fig:n2c1summaryC}
 \end{subfigure}\hspace{0.02\textwidth}
 \begin{subfigure}[t]{0.48\textwidth}
  \includegraphics[width=\textwidth]{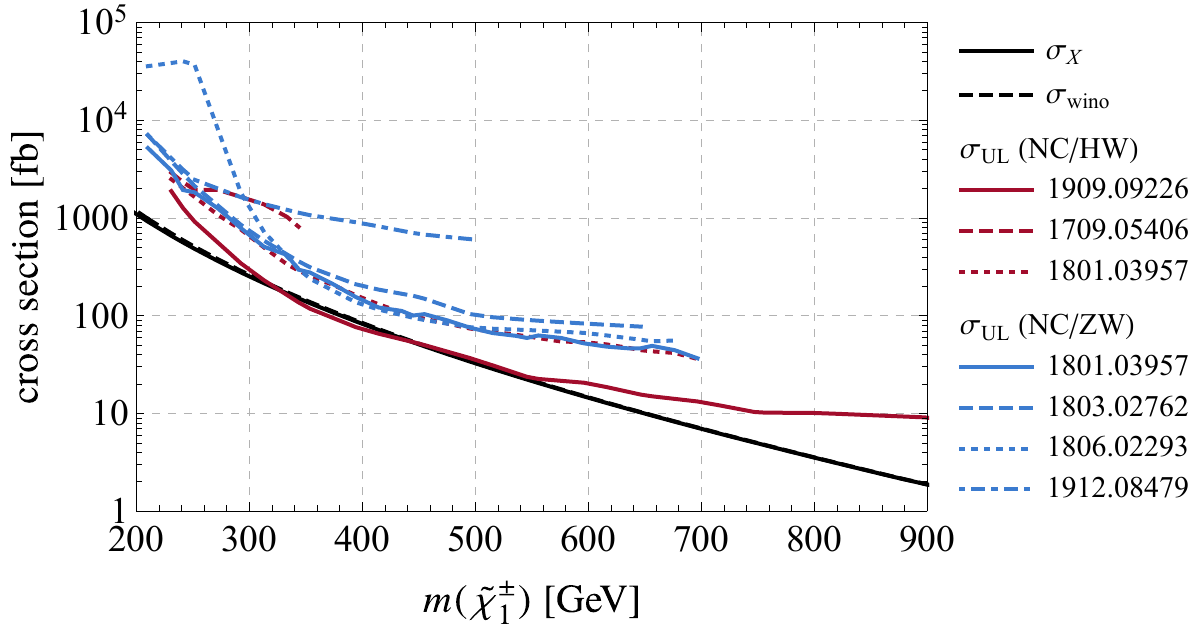}
  \caption{$\mu = 2M_2$, $m_{\tilde{\chi}^0_1}=100$\,GeV}\label{fig:n2c1summaryD}
 \end{subfigure}
\caption{
  Summary plot of the NC/HW and NC/ZW analyses. Each panel corresponds to the setup (A)--(D).
  The theoretical cross section, $\sigma_X(pp\to\neut[2]\charPM[1])$, is shown by the black solid lines, and its pure-wino limit value is shown by the black dotted lines.
  The red (blue) lines show the upper bounds on the cross section based on NC/HW (NC/ZW) analyses, which are applicable if $m_{\neut[2]}\simeq m_{\charPM[1]}$ is smaller than slepton masses.
  Three LHC results are considered in the NC/HW analyses; they are respectively based on Refs.~\cite{Aad:2019vvf}, \cite{Sirunyan:2017lae}, and \cite{Sirunyan:2018ubx} from top to bottom in the legend.
  For NC/ZW analyses, four results obtained from Refs.~\cite{Sirunyan:2018ubx}, \cite{Aaboud:2018jiw}, \cite{Aaboud:2018sua}, and \cite{Aad:2019vvi} (from top to bottom) are considered.
}
\label{fig:n2c1summary}
\end{figure}

By using Eqs.~\eqref{eq:NCHW-B} and \eqref{eq:N2C1production}, we reinterpret the ATLAS and CMS upper bounds for the process $pp\to\neut[2]\charPM[1]$.
\Figref{n2c1summary} shows the result for models in which $\neut[2]$ and $\charPM[1]$ are lighter than sleptons; otherwise constraints become much looser and irrelevant.
The black solid lines and dashed lines show the theoretical cross section and its pure-wino limit value, respectively.
The three red lines are obtained by the NC/HW analysis~\cite{Aad:2019vvf,Sirunyan:2017lae,Sirunyan:2018ubx}, while the four blue curves are from the NC/ZW process described in the next subsection.
In \figref{n2c1summaryA}, the upper bound from Ref.~\cite{Sirunyan:2018ubx} is obtained only at $m_{\charPM}=275\GeV$ and thus displayed by a marker.

As a result, the region in Fig.~\ref{fig:2m100} with $327\GeV < m_{\charPM[1]} <443\GeV$ and $m_{\charPM[1]}<m_{\tilde l\w L}$ is excluded, while no region is excluded for (A), (B), and (C).
This situation is similar to the CC/WW analysis; no regions are excluded when $M_1 = M_2/2$ because of the smaller mass separations, while in the cases with $\mu = M_2$ the higgsino component suppresses the theoretical cross sections.
Further discussions, \eg, on contributions from heavier electroweakinos, are provided in Appendix~\ref{app:EWK}.

\subsection%
[Electroweakino pair-production in final state with ZW (NC/ZW)]%
{Electroweakino pair-production in final state with $ZW$ (NC/ZW)}\label{sec:NC/ZW}
A potentially important channel of the neutralino-chargino pair-production is (\figref{NC/ZW})
\begin{equation}
p p \to \neut[2]\charPM[1] \to 
Z \tilde{\chi}^0_1 W^{\pm} \tilde{\chi}^0_1\,,
\end{equation}
resulting in the $ZW+\mET$ signature.
The search results are reported by the CMS collaboration with $\int\! dt \mathcal{L}=36\ifb$ \cite{Sirunyan:2018ubx} and the ATLAS collaboration with $\int\! dt \mathcal{L}=36\TO139\ifb$ \cite{Aaboud:2018jiw,Aaboud:2018sua,Aad:2019vvi} 
with the assumptions
$\Br(\tilde{\chi}^{0}_2 \to Z \tilde{\chi}^0_1)
=  \Br(\tilde{\chi}^{\pm}_1 \to W^{\pm} \tilde{\chi}^0_1)=1$.
We interpret the results similarly to Sec.~\ref{sec:NC/HW} but with
\begin{equation}
  \frac{B_X}{B\w{original}}=
  \Br(\tilde{\chi}^{0}_2 \to Z \tilde{\chi}^0_1)
  \Br(\tilde{\chi}^{\pm}_1 \to W^{\pm} \tilde{\chi}^0_1)
\end{equation}
and obtain upper bounds on $\sigma(pp\to\neut[2]\charPM[1])$, which are displayed by the blue lines in \figref{n2c1summary}.
It is found that no model points are excluded in \figref{summary}.
This is because $\Br (\neut[2] \to Z \neut[1])$ is at most $0.4$ in our model points and $\neut[2]$ is not likely to decay into $Z\neut[1]$ even if 
the decay channel into sleptons are forbidden.\footnote{%
The branching ratio of
$ \tilde{W}^0 \to Z \tilde{B}^0$ can be $\sim1$  when $\mathop{\mathrm{sign}}(\mu M_2) $ is negative (see Eq.~(6) in Ref.~\cite{Gori:2014oua}).
The muon $g-2$ anomaly requires $\mathop{\mathrm{sign}}(\mu M_2)$ to be positive in the present setup.}
Detail discussions, \eg, on the processes with heavier electroweakinos, are provided in Appendix~\ref{app:EWK}.

\subsection%
[Electroweakino pair-production in final state with three leptons (NC/3L)]%
{Electroweakino pair-production in final state with three leptons (NC/3L)}\label{sec:NC/3L}

Electroweakino pair-production $pp\to\neut[2]\charPM[1]$ is severely constrained, if sleptons are lighter than the electroweakinos, because
the electroweakinos decay via the sleptons to provide three leptons in the final state (\figref{NC/3L}):
\begin{equation}
pp\to \neut[2]\charPM[1]\to 
\begin{cases}
 (l \tilde l\w L)(\nu \tilde{l}\w L)
 \to
 (ll \neut[1])(\nu l \neut[1])\,,
 \\
  (l \tilde{l}\w L) (l \tilde{\nu})
 \to
 (l l \neut[1])(l \nu \neut[1])\,.
\end{cases}
\end{equation}
This process, NC/3L, is searched for in, \eg, its $3\ell+\mET$ or same-sign $2\ell$ plus $\mET$ signatures by the CMS and ATLAS collaborations.
The latest bounds with $\int\! dt \mathcal{L}=36\ifb$ are provided in Refs.~\cite{Sirunyan:2017lae,Aaboud:2018jiw}, where a simplified model with branching ratios
\begin{equation}
\begin{split}
  & \Br (\neut[2]\to \tilde e\w L  \bar e\text{~or~}\tilde e\w L ^\ast e )
 = \Br (\neut[2]\to \tilde \mu\w L  \bar \mu\text{~or~} \tilde \mu\w L ^\ast \mu ) 
 = \Br (\neut[2]\to \tilde \tau\w L  \bar \tau\text{~or~} \tilde \tau\w L ^\ast \tau ) = 1/6\,,
\\
 & \Br (\charP[1] \to  \tilde e\w L ^{\ast}  \nu_{e}\text{~or~} \tilde \nu \bar e) 
 = \Br (\charP[1] \to  \tilde \mu\w L ^{\ast}  \nu_{\mu}\text{~or~} \tilde \nu \bar \mu) 
 = \Br (\charP[1] \to  \tilde \tau\w L ^{\ast}  \nu_{\tau}\text{~or~} \tilde \nu \bar \tau) = 1/3\,,
\end{split}
 \label{eq:NC3L-expassum}
\end{equation}
is considered.
Their results can be reinterpreted with
\begin{equation}
\begin{split}
 \frac{B_{X}}{B\w{original} }&=\frac{1}{0.273} 
 \left[ \Br (\neut[2]\to \tilde \ell\w L  \bar \ell, \tilde \ell\w L ^\ast \ell ) +
 \frac{3}{4} p_{\tau \to \ell}^2\Br (\neut[2]\to \tilde \tau\w L  \bar \tau, \tilde \tau\w L ^\ast \tau )\right]
 \\
 &\times \left[ \Br (\charPM[1] \to  \tilde \ell\w L ^{\ast}  \nu_{\ell}, \tilde \nu \bar \ell, \tilde \ell\w L \bar \nu_{\ell}, \tilde \nu ^{\ast} \ell)
 +
 p_{\tau \to \ell} \Br (\charPM[1] \to  \tilde \tau\w L ^{\ast}  \nu_{\tau}, \tilde \nu \bar \tau,  \tilde \tau\w L \bar \nu_{\tau}, \tilde \nu ^{\ast} \tau)
 \right]\,.
\end{split}
\label{eq:AA3L}
\end{equation}
When the final state includes $\tau$ leptons, their contributions are taken into account via leptonic $\tau$ decays. 
Here, $p_{\tau \to \ell} \equiv  \Br(\tau \to e \bar{\nu}_{e} \nu_{\tau}(\gamma))+ \Br(\tau \to \mu \bar{\nu}_{\mu} \nu_{\tau}(\gamma)) = 0.352$ \cite{Tanabashi:2018oca}, and
the factor $3/4$ takes care of the opposite-sign same-flavor leptons. 
The overall normalization $0.273$ is determined such that the right-hand side of Eq.~\eqref{eq:AA3L} becomes unity for the branching ratios adopted by the ATLAS and CMS collaborations.
On the other hand, the production cross section, 
$\sigma(p p\to \tilde{\chi}^0_2 \tilde{\chi}^{\pm}_1)$, is estimated in the same way as Eq.~\eqref{eq:N2C1production},
and $\sigma\UL(p p\to \tilde{\chi}^{0}_2 \tilde{\chi}^{\pm}_1)$ is available in 
Refs.~\cite{10.17182/hepdata.81996.v1/t80,Sirunyan:2017lae}.

An important point in the LHC analyses is that both of the ATLAS and CMS collaborations assume specific mass spectra of electroweakinos and sleptons, which determine the lepton energies.
Defining a mass difference ratio,
\begin{equation}
 x=\frac{m_{\smuL} - m_{\neut[1]}}{m_{\charPM[1]} - m_{\neut[1]}}\,,
 \label{eq:x}
\end{equation}
the CMS collaboration considers three different mass spectra, $x = (0.05,\,0.5,\,0.95)$, while the ATLAS studies $x=0.5$.
Following their analyses, we investigate the constraints on the same mass spectra,  $x = (0.05,\,0.5,\,0.95)$. The corresponding model points are displayed by the dashed black lines in Fig.~\ref{fig:summary}.

Based on Eq.~\eqref{eq:bound} with Eqs.~\eqref{eq:N2C1production} and \eqref{eq:AA3L}, we obtain the upper bounds $\sigma\ULX$, which are shown in \figref{nc3lsummary} by the red and blue lines.
Two lines are drawn for each analysis.
The lower line is obtained by the above-described (``standard'') method, while the upper line shows the conservative bound, which we explain in the next paragraph; the split shows uncertainty on our analysis.
Consequently, the excluded regions on each $x$ are obtained as shown by the magenta bars on dashed black lines in Fig.~\ref{fig:summary}; the thicker (thinner) bars correspond to the conservative (``standard'') exclusion. %
As the CMS and ATLAS collaborations report similar bounds for $x=0.5$, we use the stronger one in \figref{summary} for simplicity.

\begin{figure}[t]
 \renewcommand\thesubfigure{\Alph{subfigure}}
 \centering
 \begin{subfigure}[t]{0.48\textwidth}
  \includegraphics[width=\textwidth]{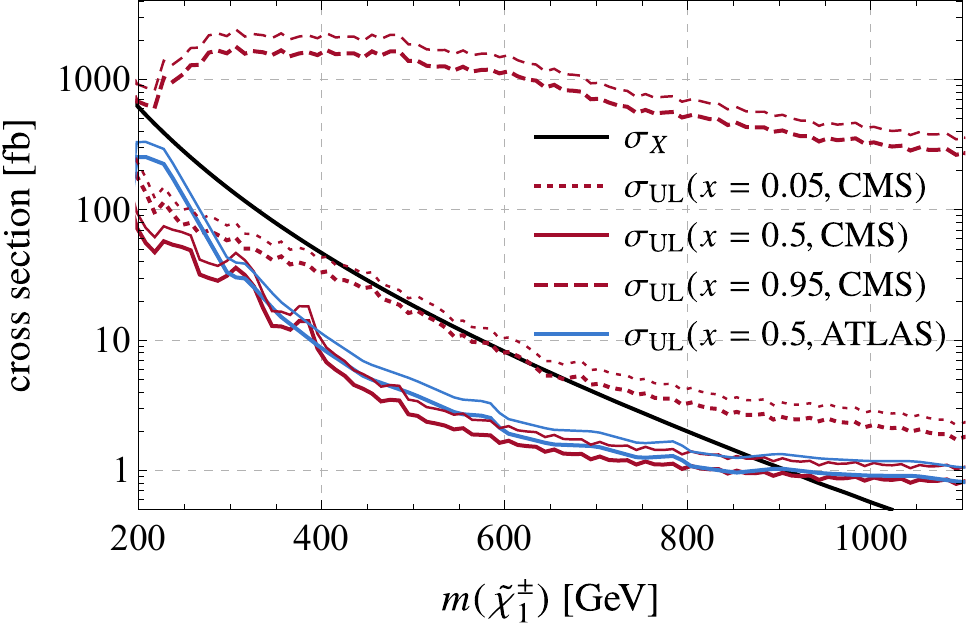}
  \caption{$\mu = M_2$, $M_1=M_2/2$}\label{fig:nc3lsummaryA}
 \end{subfigure}\hspace{0.02\textwidth}
 \begin{subfigure}[t]{0.48\textwidth}
  \includegraphics[width=\textwidth]{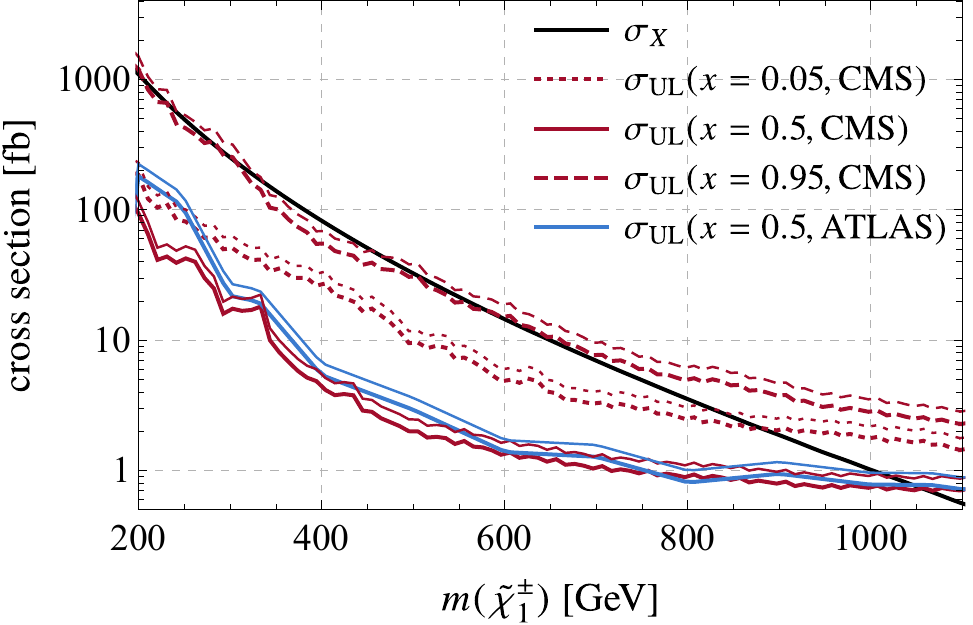}
  \caption{$\mu = 2M_2$, $M_1=M_2/2$}\label{fig:nc3lsummaryB}
 \end{subfigure}
\par\vspace{1em}
 \begin{subfigure}[t]{0.48\textwidth}
  \includegraphics[width=\textwidth]{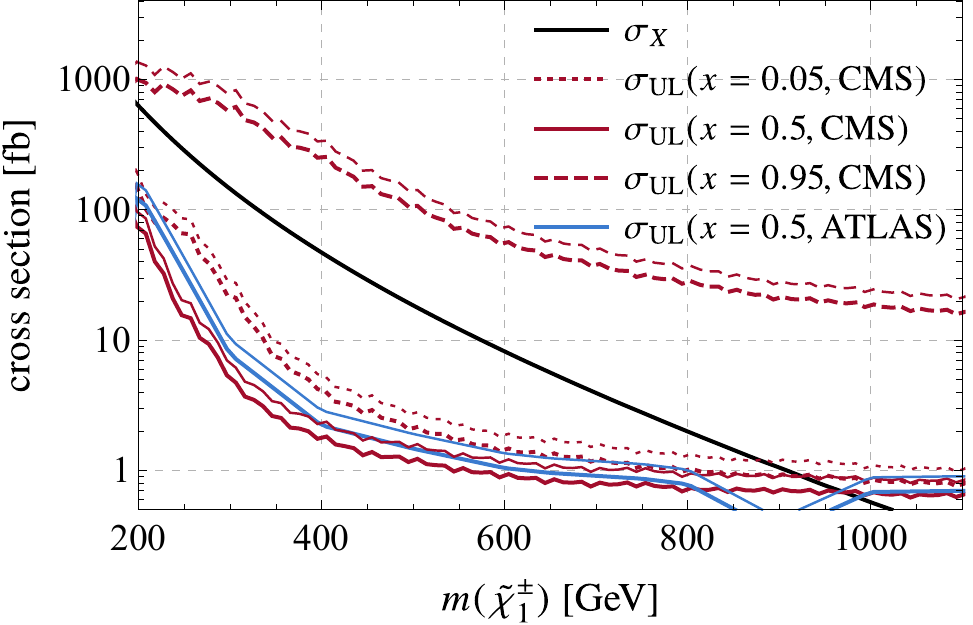}
  \caption{$\mu = M_2$, $m_{\tilde{\chi}^0_1}=100$\,GeV}\label{fig:nc3lsummaryC}
 \end{subfigure}\hspace{0.02\textwidth}
 \begin{subfigure}[t]{0.48\textwidth}
  \includegraphics[width=\textwidth]{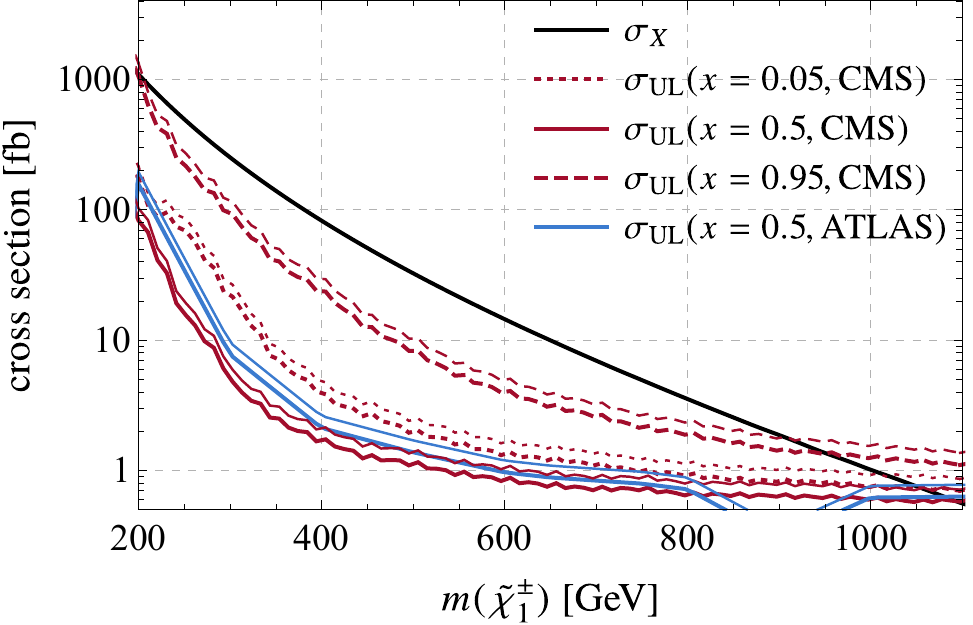}
  \caption{$\mu = 2M_2$, $m_{\tilde{\chi}^0_1}=100$\,GeV}\label{fig:nc3lsummaryD}
 \end{subfigure}
\caption{
  Summary plot of the NC/3L analysis.
  The theoretical cross section, $\sigma_X(pp\to\neut[2]\charPM[1])$, is common to all the $x$s and shown by the black solid line.
  By reinterpretation of the CMS result~\cite{Sirunyan:2017lae}, we obtain upper bounds on the cross section for the model points with $x=0.05$, $0.5$, and $0.95$, which are respectively drawn by the red dotted, red solid, and red dashed lines.
  Reinterpretation of the ATLAS result~\cite{Aaboud:2018jiw,10.17182/hepdata.81996.v1/t80} provides upper bounds for $x=0.5$ drawn by the blue solid lines.
  Each of the upper bounds is evaluated in two criteria and thus drawn by two lines; the lower line corresponds to the ``standard'' limit while the upper one is the conservative limit; see text for details.
}
\label{fig:nc3lsummary}
\end{figure}

Our results are based on the LHC analyses under the flavor-universal assumptions, Eq.~\eqref{eq:NC3L-expassum}.
The CMS collaboration also studied a model in which the chargino decays exclusively into $\tilde\tau\bar\nu$~\cite{Sirunyan:2017lae}.
As a crosscheck, we calculated the cross-section upper bounds for this mode, using our reinterpretation method, Eqs.~\eqref{eq:bound} and \eqref{eq:AA3L}, and obtained a stronger bound than the CMS.
This inconsistency should be regarded as an uncertainty in our analysis and we consider that it mainly originates in an overestimation of the acceptance of leptons from taus.
Namely, we in Eq.~\eqref{eq:AA3L} assume that leptons from tau decays are detected with the same efficiency as those from SUSY particles, but in fact the former ones have smaller energy and thus the efficiency should be smaller.
To understand the influence of this uncertainty, we also evaluate the cross-section upper bounds with discarding the tau-originating leptons, \ie, with Eq.~\eqref{eq:NC3L-expassum} but $p_{\tau\to\ell}$ substituted by zero, and obtain conservative bounds, shown by the upper lines in \figref{nc3lsummary}.
We found that the conservative bounds are weaker than the ``standard'' ones by $25\text{--}35\%$ as shown in \figref{nc3lsummary}, and that the results do not change much.
For example, for $x=0.5$, the end points of the magenta lines in Fig.~\ref{fig:summary} are shifted only by ${\cal O}(10)$\,GeV.

We observe in \figref{summary} that this NC/3L channel provides stringent constraints on the model points with $m_{\tilde\mu\w L}<m_{\charPM[1]}$, as we have found in the previous work~\cite{Endo:2013bba}.
The only exception is for $x=0.95$ with $\mu=M_2$ (\figsref{1gut}{1m100}).
For $x=0.5$ or $0.05$, the electroweakinos $\neut[2]$ and $\charPM[1]$ mainly decay into $\tilde l\w L$ or $\tilde\nu$.
Meanwhile, for $x=0.95$, the degeneracy of sleptons with $\neut[2]$ and $\charPM[1]$ suppresses the decays of $\neut[2]$ and $\charPM[1]$ into sleptons.
As a result, the decay channels $\neut[2]\to h\neut[1]$ and $\charPM[1]\to W^\pm\neut[1]$ tend to be dominant for our model points with $x=0.95$ and $\mu=M_2$ and no constraints are obtained by this NC/3L analysis.
Note that those model points are not excluded by the NC/HW analysis either.
Meanwhile, in the model points with $x=0.95$ and $\mu=2M_2$, the decays $\neut[2]\to Z\neut[1],\, h\neut[1]$ are suppressed by the neutralino mixings and $\neut[2]$ tends to decay into $\tilde\nu$s, which are slightly lighter than $\tilde l\w L$.

\subsection{Results}

All the relevant LHC bounds on the SUSY parameter space for the muon $g-2$ anomaly are summarized in Fig.~\ref{fig:summary}.
The bounds from the SLSL, CC/WW, NC/HW, and NC/3L analyses are shown by the blue-shaded region, red-shaded region with the dashed-dotted boundary, red-shaded region with the dashed boundary, and magenta bar, respectively.
It is noticed that we have not investigated the NC/3L bounds for arbitrary $x$ value in Eq.~\eqref{eq:x}.
According to the blue-dotted lines in Fig.~1 of Ref.~\cite{Endo:2013bba}, it is expected that the bounds on $x=0.05,\,0.5,\,0.95$ are continuously connected with a peak around $x=0.5$.

For $m_{\tilde{\mu}\w L} \lesssim m_{\charPM[1]}$, \ie, the region below the thick black line, it is found that the regions favored by $\Delta\amu$ are severely constrained by the SLSL and NC/3L searches.
On the other hand, for $m_{\tilde{\mu}\w L} > m_{\charPM[1]}$, \ie, the region above the thick black line,
it is found that the LHC bounds are drastically loosened.
The SLSL search excludes regions with $m_{\tilde{\mu}\w L} \gtrsim m_{\charPM[1]}$, and it is the only bound in the cases of (A) and (B).
In the cases of (C) and (D), the CC/WW and NC/HW searches become relevant.
The bound is very loose in (C) because of the smaller production cross section of the electroweakinos.
A wide excluded regions appear only when the electroweakinos produced are wino-like and the mass splitting between them and the LSP is sufficient, namely, in the case (D).

\section{Conclusions and Discussion}
\label{sec:conclusion}

In this paper, we revisited the LHC bounds on the chargino-dominated SUSY solution to the muon $g-2$ anomaly.
The latest bounds from the LHC Run~2 are investigated for electroweakinos and sleptons.
It was found that the model parameter regions are severely constrained when $m_{\tilde{\mu}\w L} \lesssim m_{\tilde{\chi}^{\pm}_1}$, while a wide parameter region still survives for $m_{\tilde{\mu}\w L} \gtrsim m_{\tilde{\chi}^{\pm}_1}$.
Interpreting the four panels of \figref{summary}, we conclude that models with $m_{\tilde{\mu}\w L} \lesssim m_{\tilde{\chi}^{\pm}_1}$ are strongly disfavored as the explanation for $\Delta\amu$ as far as they satisfy $0<M_1/M_2<1/2$,
$1\le\mu/M_2 \le2$, $m\w R\gtrsim1\TeV$, and $\tan\beta\le40$.

In this paper, we focused on $\mu/M_2 \geq 1$.
Meanwhile, if higgsinos are lighter than winos, $\neut[2]$ and $\charPM[1]$ are mostly composed of the higgsinos.
Then, models with $m_{\tilde l\w L}>m_{\charPM[1]}$ receive more severe constraints from the SLSL search because $\Br (\tilde \ell\w L \to \nu_{\ell} \charM[1])$ is suppressed by the lepton Yukawa coupling, and thus $\Br (\tilde \ell\w L \to {\ell} \neut[1])$ is amplified.
Meanwhile, the CC/WW and NC/HW bounds from $\neut[2]$ and $\charPM[1]$ are loosened for such model points because the electroweakino production cross section, $\sigma (p p \to \neut[2] \charPM[1] )$, is likely to be suppressed (see Appendix~\ref{sec:AppA1}).
The NC/ZW bound from $\neut[2]\charPM[1]$ may be relevant because the branching ratio of $\neut[2] \to Z \neut[1] $ is enhanced (see Appendix~\ref{app:EWK}).
Also, contributions from the heavier electroweakinos, which are now wino-like, are expected to be restrictive.
Analysis for models with $\mu/M_2<1$ and $m_{\tilde l\w L}<m_{\charPM[1]}$ are more complicated.
Firstly, because $\neut[2]$ and $\charPM[1]$ are now higgsino-like and the muon $g-2$ anomaly motivates large $\tan\beta$, they tend to decay into $\tilde\tau$ or $\tilde \nu_\tau$ as far as the channel do not experience phase-space suppression.
Thus, searches for the two or more tau-leptons plus $\mET$ signature are expected to be sensitive to such models. Further discussion is found in, \eg, Ref.~\cite{Endo:2017zrj}.
In addition, the heavier electroweakinos, especially wino-like ones, become relevant for these models because of their larger production cross sections.
They provide cascade-decay signatures, \eg, $\charP[2]\to W^+\neut[2]\to W^+Z\neut[1]$, or may result in the NC/3L decay process.
Dedicated Monte Carlo analyses are necessary to study the signatures.

One may think of increasing $\tan\beta$, which we fix $\tan \beta = 40$ as a reference value.
The SUSY contributions to the muon $g-2$ are almost proportional to $\tan \beta$;
$a_{\mu}^{\rm SUSY}$ is amplified, \eg, by $50\,\%$ for $\tan \beta = 60$. 
On the other hand, the LHC bounds are less sensitive to $\tan \beta$ as long as $\tan\beta\gtrsim 10$ and the above conclusion does not change.

In the cases (A) and (B), the quasi-GUT relation $M_1:M_2=1:2$ was considered, while the gluino $\tilde g$ was assumed to be decoupled.
If we include the gluino and use the GUT relation $M_1:M_2:M_3\simeq1:2:6$, a large portion of the parameter space is constrained by the gluino searches.
According to LHC Run~2 results \cite{Aaboud:2017vwy,ATLAS:2019vcq}, gluino is excluded up to $m_{\tilde g}\sim 2\TeV$, which implies the region of $m_{\charPM[1]} \lesssim 600\,$GeV in Fig.~\ref{fig:summary} would be excluded, although decisive bounds strongly depend on squark mass spectra as well as the gluino branching ratios.

In our analysis, we have focused on the parameter regions where the LSP is the lightest neutralino. In Fig.~\ref{fig:summary}, near the boundary of the gray-filled region (around the black solid line of $x=0.05$), the thermal relic abundance of the LSP can explain the observed dark matter abundance by the coannihilation between the LSP neutralino and the slepton. However, in the current setup, the LHC constraints already excluded most of the regions that can simultaneously explain the thermal relic abundance and the muon $g-2$, and part of those regions are also constrained by the direct detection experiments (cf.~Refs.~\cite{Endo:2017zrj,Chakraborti:2017vxz,Ajaib:2017zba,Belyaev:2018vkl,Abel:2018ekz,Abdughani:2019wai}).

Let us compare the results with our previous result, namely, \figsref{1gut}{2gut} with Fig.~1 (a) and (b) of Ref.~\cite{Endo:2013bba}.
Two differences in the model set-up should be in mind; firstly, the previous work includes models with $M_1:M_2:M_3=1:2:6$ in its scope, and thus bounds from gluino searches, which we called  ``J-search'', were drawn.
The corresponding bounds are not provided in this work, just briefly discussed in this section, as we concentrate on $M_3\gg M_2$.
The other bounds, once we called ``L-search'' bounds and drew by the blue curve, correspond to the NC/3L bound in this work.
However, here comes the second difference: the previous work assumes staus are decoupled, while here $\tilde\tau\w L$ and $\tilde\nu_{\tau}$ have similar mass as $\tilde\ell\w L$ and $\tilde \nu_{\ell}$.
The ``L-search'' results should not be directly compared to the NC/3L results because the new setup allows electroweakinos to decay into staus as well; if we did our analysis under the previous setup, it would give $B_X/B\w{original}\approx0.5/0.273$ according to Eq.~\eqref{eq:AA3L}, and the resulting $\sigma\ULX$ would be stronger by a factor $\sim1.8$.
Keeping these two differences in mind, and noting that the previous (new) figures are drawn with the soft mass parameter (the physical masses) as the axes, we find that the LHC Run~2 has provided surprisingly good sensitivity to the parameter space motivated by the muon $g-2$ anomaly.

Finally, we briefly comment on the future prospects. 
According to Refs.~\cite{ATL-PHYS-PUB-2018-048,CidVidal:2018eel}, the HL-LHC may test the region with $m_{\charPM[1]} \lesssim 1.2\text{--}1.3\TeV$ of a simplified model by analyzing the NC/HW channel.
At a future $100\TeV$  $pp$ collider, the parameter region with
$m_{\charPM[1]} \lesssim 1.4\,(3.4)\TeV$
 can be tested when $M_1 = M_2/2$ ($m_{\neut[1]}=100$\,GeV) is taken
 \cite{Gori:2014oua,Bramante:2014tba}.
Indirect searches at future colliders also probe the electroweakinos \cite{Matsumoto:2017vfu,Matsumoto:2018ioi,Chigusa:2018vxz,Abe:2019egv};
for example, a $100\TeV$ $pp$ collider has potential to exclude charginos with $m_{\charPM[1]} \lesssim 1.7\text{--}2.3\TeV$.

If the muon $g-2$ anomaly is strengthened and confirmed by the future experimental and theoretical studies, it will be a strong motivation for the BSM physics at around the electroweak scale, including the low energy SUSY. We hope that this study will be a useful step for further studies and the model building in such a scenario.

\section*{Acknowledgments}
This work is supported in part by the Grant-in-Aid for 
Innovative Areas (No.19H04612 [KH], No.19H05810 [KH], No.19H05802 [KH]),
Scientific Research A (No.16H02189 [KH]), 
Scientific Research B (No.16H03991 [ME]), and 
Early-Career Scientists (No.16K17681 [ME] and No.19K14706 [TK]).
The work of T.K.\ is also supported by 
the Israel Science Foundation (Grant No.~751/19).
The work of S.I.\ is partially supported by the MIUR-PRIN project 2015P5SBHT 003 ``Search for the Fundamental Laws and Constituents''.
Diagrams in Fig.~\ref{fig:diagram} are drawn with \package[]{TikZ-Feynman} \cite{Ellis:2016jkw}.

\appendix

\section{Auxiliary materials}
\label{app:Auxiliary}
In this appendix we provide extra materials and additional discussions.

\subsection{Electroweakino production cross section in pure-ino limits}
\label{sec:AppA1}

\begin{figure}[t]
 \centering
 \includegraphics[height=150pt]{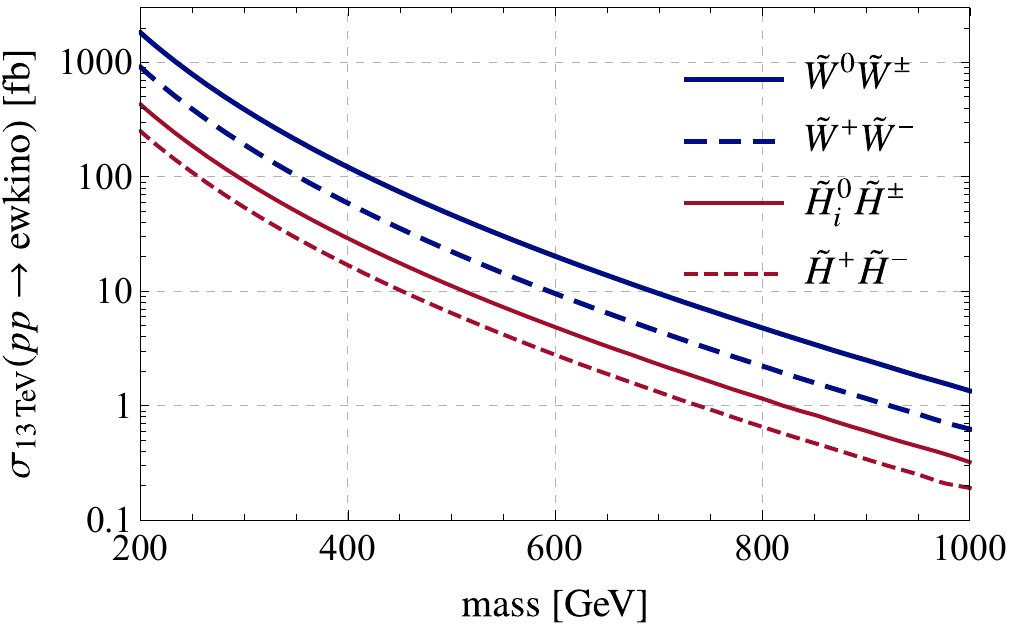}
 \caption{
 \label{fig:pure-ino-xs}
 Pure electroweakino pair-production cross sections at the NLO-NLL accuracy at the LHC with $\sqrt{s}=13\TeV$, where  squarks are decoupled and only the Drell-Yan processes contribute.
 Masses of the produced electroweakinos are assumed to be degenerate.
 Contributions from $\tilde\chi^+$ and $\tilde\chi^-$ are summed in the neutralino-chargino pair-productions.
}
\end{figure}

\Figref{pure-ino-xs} illustrates the electroweakino pair-production cross section at the LHC with $\sqrt{s}=13\TeV$ in the pure-wino and pure-higgsino limits.
Here, squarks are assumed to be decoupled, and only the Drell-Yan processes, \ie, diagrams with $s$-channel exchange of SM gauge bosons, contribute to the production at the leading order.
The values are calculated at the NLO-NLL accuracy~\cite{Beenakker:1999xh,Debove:2010kf,Fuks:2012qx,Fuks:2013vua,Fiaschi:2018hgm} and obtained from Ref.~\cite{LHCSUSYCSWG}.
For the pure-higgsino limit, in which two higgsinos compose two degenerate neutralinos denoted by $\tilde H_i^0$ ($i=1,2$), the neutralino-chargino production cross section is effectively twice as large as the red solid line.
It is shown that
\begin{equation}
\sigma(pp \to \tilde{W}^0 \tilde{W}^{\pm}) > 
\sigma(pp \to \tilde{W}^+ \tilde{W}^{-}) > 
\sigma(pp \to \tilde{H}^0_i \tilde{H}^{\pm}) >
\sigma(pp \to \tilde{H}^+ \tilde{H}^{-})\,,
\end{equation}
where each of the differences is around a factor of two.

\subsection{Auxiliary plots for the electroweakino analyses}\label{app:EWK}

\def\fourfigs#1{
  \begin{minipage}{0.242\textwidth}\hfill\includegraphics[height=140pt]{#1_1gut.pdf}\end{minipage}
  \begin{minipage}{0.242\textwidth}\hfill\includegraphics[height=140pt]{#1_2gut.pdf}\end{minipage}
  \begin{minipage}{0.242\textwidth}\hfill\includegraphics[height=140pt]{#1_1m100.pdf}\end{minipage}
  \begin{minipage}{0.242\textwidth}\hfill\includegraphics[height=140pt]{#1_2m100.pdf}\end{minipage}
}

\begin{figure}[p]
 \centering \fourfigs{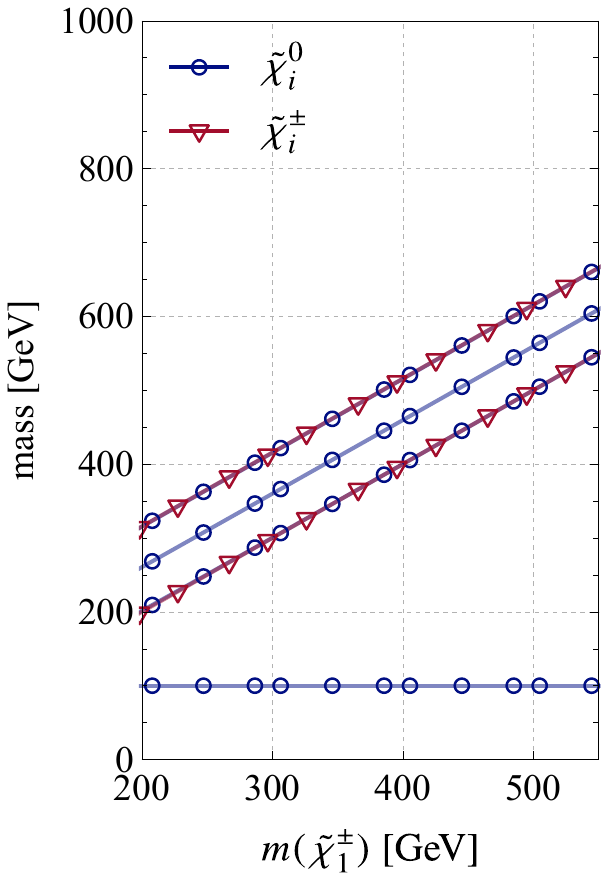}
\caption{The mass spectra of electroweakinos ($\neut[\text{1--4}]$ and $\charPM[\text{1--2}]$) in (A), (B), (C), and (D) in Eq.~\eqref{eq:parameterspace} from left to right, respectively.}
\label{fig:n2c1mass}
\end{figure}

\begin{figure}[p]
 \centering \fourfigs{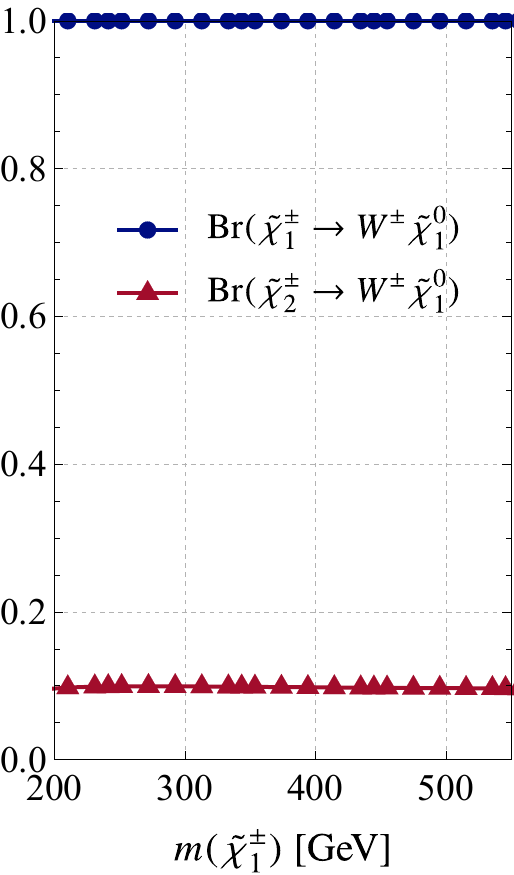}
\caption{Chargino branching ratios into $W^\pm\neut[1]$ for model points in which the chargino is lighter than sleptons; otherwise, the branching ratio is smaller than displayed. As in \figref{n2c1mass}, panels respectively correspond to (A), (B), (C), and (D) from left to right.}
\label{fig:n2c1chardecay}
\end{figure}

\begin{figure}[p]
 \centering \fourfigs{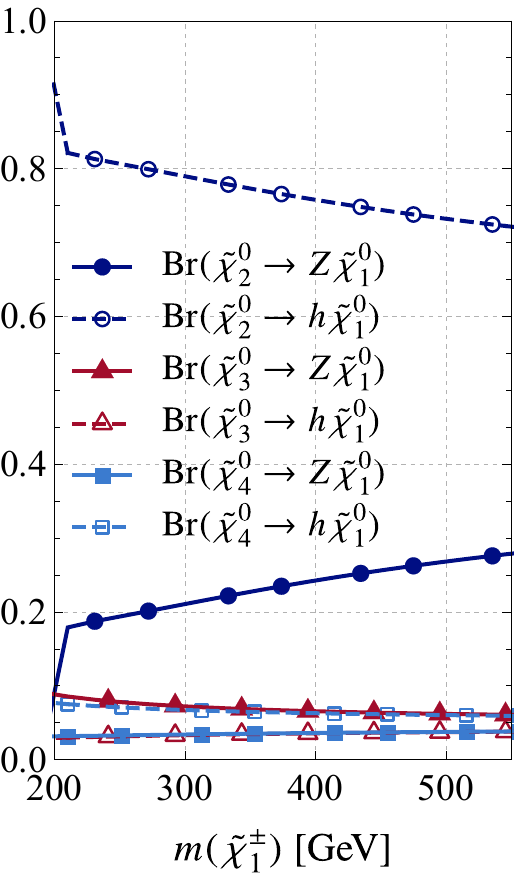}
\caption{Neutralino branching ratios for model points in which the neutralino is lighter than sleptons; otherwise, the branching ratio is smaller than displayed. As in \figsref{n2c1mass}{n2c1chardecay}, panels respectively correspond to (A), (B), (C), and (D) from left to right.}
\label{fig:n2c1neutdecay}
\end{figure}

We provide auxiliary materials for analyses of the electroweakino productions, CC/WW, NC/HW, and NC/ZW.
The masses of electroweakinos are shown in \figref{n2c1mass}. 
The four panels respectively correspond to, from left to right, each of the setups (A), (B), (C), and (D) in Eq.~\eqref{eq:parameterspace} and \figref{summary}; in particular, the first two panels are under the quasi-GUT assumption $M_1=M_2/2$, while the latter two are with $m_{\neut[1]}=100\GeV$.
Since we use the masses and mixings calculated at the tree level, they are independent of the slepton masses (the vertical axis of \figref{summary}).

\Figref{n2c1chardecay} shows the branching ratios of $\charPM[i]$ into $W^\pm\neut[1]$, while the branching ratios of $\neut[i]$ are shown in \figref{n2c1neutdecay}.
All the figures are drawn with the assumption that each of the electroweakinos are lighter than the sleptons ($\tilde l\w L$ and $\tilde\nu$); otherwise it may decay into the sleptons and its branching ratios into SM bosons become much smaller than the plotted values.

\begin{figure}[p]
 \centering \fourfigs{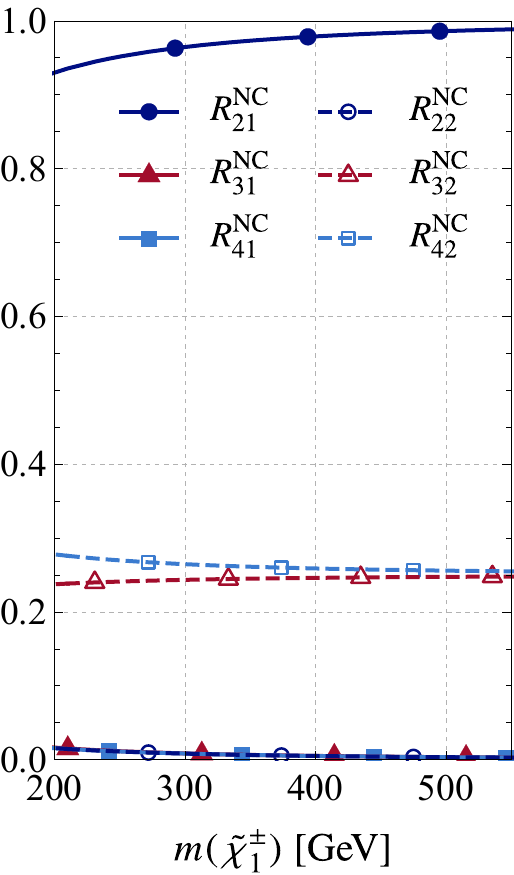}
 \caption{The cross section reduction factor $R^{\mathrm{NC}}_{ij}$ for our model points; panels respectively correspond to (A), (B), (C), and (D) from left to right.}
\label{fig:n2c1cmean}
\end{figure}

Let us revisit the result of NC/HW and NC/ZW analyses, namely \figref{n2c1summary}.
Both analyses are relevant because $\neut[2]$ has two sizable decay channels, $Z\neut[1]$ and $h\neut[1]$, when its decays into sleptons are kinematically forbidden.
These branching ratios form the $B$-ratio in Eq.~\eqref{eq:ULtranslation} as
\begin{equation}
 \left(\frac{B_X}{B\w{original}}\right)_{ij} =
\begin{cases}
   \Br(\neut[i]\to h\neut[1])\Br(\charPM[j]\to W^\pm\neut[1]) & \text{for NC/HW},\\
   \Br(\neut[i]\to Z\neut[1])\Br(\charPM[j]\to W^\pm\neut[1]) & \text{for NC/ZW},
\end{cases}
\end{equation}
where we extend the definition to general process $pp\to\neut[i]\charPM[j]$.
Since this factor is smaller than one, the cross section upper bound $\sigma\ULX$ becomes larger than its original value, $\sigma\ULorig$, as stated in Eq.~\eqref{eq:bound}.
Meanwhile, in Eq.~\eqref{eq:Xsecchi21}, we approximately give the theoretical cross section by $\sigma_X\approx R^{\mathrm{NC}}_{ij}\sigma\w{wino}$.
It means, in \figref{n2c1summary}, the cross section $\sigma_X$ is obtained by shifting the $\sigma\w{wino}$ line downwards by the factor $R^{\mathrm{NC}}_{21}$.
Figure~\ref{fig:n2c1cmean} shows the cross section reduction factors $R^{\mathrm NC}_{ij}$ for our model points.
Then, in similar manner, one may consider that the curves of $\sigma\ULX$ are obtained by shifting $\sigma\ULorig$ upwards by the factor $({B\w{original}}/{B_X})_{21}$.
Here, the lines for $\sigma\ULorig$ will be the same in all the four panels if drawn.

\begin{figure}[p]
 \centering \fourfigs{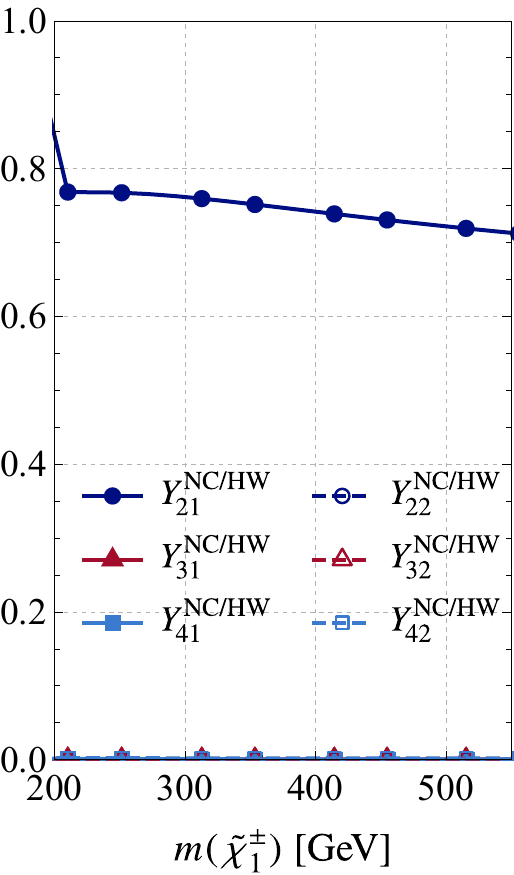}
\caption{The effective yield, defined in Eq.~\eqref{eq:yield}, of the NC/HW analysis of $pp\to\neut[i]\charPM[j]$ production. Each panel, from left to right, corresponds to the parameter set (A)--(D) in Eq.~\eqref{eq:parameterspace} and \figref{summary}.
The values are for the model points with $\neut[i]$ and $\charPM[j]$ heavier than the sleptons; otherwise they are significantly smaller.}
\label{fig:n2c1Y-HW}
\end{figure}

\begin{figure}[p]
 \centering \fourfigs{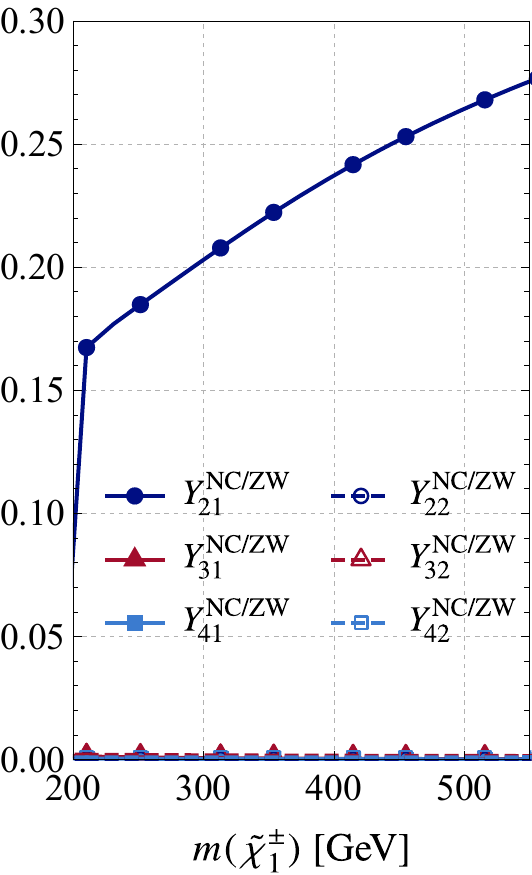}
\caption{The same as \figref{n2c1Y-HW} but of the NC/ZW analysis.}
\label{fig:n2c1Y-ZW}
\end{figure}

It is then found useful to introduce the effective yield,
\begin{equation}
 Y_{ij} \equiv R_{ij}^{\mathrm{NC}} \cdot \left(\frac{B_X}{B\w{original}}\right)_{ij}
\label{eq:yield}
\end{equation}
for each of NC/HW and NC/ZW, and for all the possible production channels.
Their numerical values are shown in \figsref{n2c1Y-HW}{n2c1Y-ZW}.
With $Y_{ij}$, Eq.~\eqref{eq:bound} is rewritten as
\begin{equation}
Y_{ij}\cdot
\sigma_{\text{NLO-NLL}}(pp\to\tilde W^0\tilde W^\pm)
>\sigma\ULorig\,.
\label{eq:rewrittencondition}
\end{equation}
This implies that the original bounds are loosened by the two factors, $R^{\mathrm{NC}}_{ij}$ coming from the cross section reduction and $(B_X/B\w{original})_{ij}$ coming from the branching ratio, which are smaller than one in our model points.
Notably, $Y_{21}^{\text{NC/HW}}$ is significantly larger than $Y_{21}^{\text{NC/ZW}}$ for all the model points because $\neut[2]$ mainly decays into $h\neut[1]$.
Therefore, although searches for the $ZW^\pm+\mET$ signature tend to give smaller $\sigma\ULorig$ than those for $hW^\pm+\mET$, an exclusion is obtained only in the NC/HW analysis.

Let us now briefly discuss the contributions from heavier electroweakinos, \ie, $pp\to\neut[i]\charPM[j]$ with $(i,j)\neq(2,1)$.\footnote{For the case $(i,j)=(1,1)$, the production cross section is likely to be suppressed.} 
From \figsref{n2c1Y-HW}{n2c1Y-ZW}, we find that such contributions are negligible in NC/HW; it is sufficient to consider $pp\to\neut[3]\charPM[1]$ in NC/ZW, and in the cases with $M_2=\mu$.
For the case (C), however, the $\neut[3]\charPM[1]$ contribution is at most comparable to the $\neut[2]\charPM[1]$ contribution, and in fact less because $\neut[3]$ is heavier than $\neut[2]$. Therefore, from \figref{n2c1summaryC}, it is observed that no region will be excluded even if we consider the $\neut[3]\charPM[1]$ contribution.

In the case (A), in fact, the NC/ZW signature is mainly provided by $\neut[3]\charPM[1]$, not by $\neut[2]\charPM[1]$,
because the branching ratio $\Br(\neut[3]\to Z\neut[1])$ is significantly larger than that of $\neut[2]$ enough to compensate $R^{\mathrm{NC}}_{31}$, which is smaller than $R^{\mathrm{NC}}_{21}$ because of the higgsino component in $\neut[3]$.
Consequently, $\neut[3]\charPM[1]$ process will provide the best sensitivity for the NC/ZW analysis in this case (A).
The full analysis requires Monte Carlo simulation because in the LHC results, $\sigma\ULorig$ is reported with an working assumption that the neutralino and chargino have a common mass; the $\neut[3]\charPM[1]$ pair has mass difference of $\sim50\GeV$ (see \figref{n2c1mass}) and its collider signature will be different from those with a common mass.
Nevertheless, we may argue that this NC/ZW signature from $\neut[3]\charPM[1]$ pair would not exclude any parameter space in \figref{1gut}.
From \figref{n2c1summaryA}, the ratio $\sigma\ULX/\sigma_X$ is larger than 10; this gap will be smaller in the $\neut[3]\charPM[1]$ case but only by $Y_{31}^{\mathrm{NC/ZW}}/Y_{21}^{\mathrm{NC/ZW}}\approx2$, which is far insufficient to have a exclusion region.
One may also check that the larger mass difference $m_{\neut[3]}-m_{\neut[1]}$ does not considerably improve this situation.
Therefore, we conclude that $\neut[3]\charPM[1]$ does not provide additional exclusion region even in \figref{1gut}, which will stay even with full $139\ifb$ results.

\section{Neutralino-chargino production cross section}
\label{app:xs}
In this appendix, we discuss the cross section ratio
\begin{equation}
\frac
  {\sigma_{\text{tree}}(pp\to\neut[i]\charPM[j])}
  {\sigma_{\text{tree}}(pp\to\neut[i]\charPM[j])|\w{wino}}
\end{equation}
under the assumption that all squarks are decoupled.
Here, $\neut[{1\TO4}]$ and $\charPM[1,2]$ are mass eigenstates of neutralinos and charginos, respectively.
Also, ${\sigma_{\text{tree}}(pp\to\neut[i]\charPM[j])|\w{wino}}$ is the tree-level cross section in the wino-like limit, which will be defined below.
If $m_{\neut[i]}=m_{\charPM[j]}$, it equals to the pure-wino production cross section
\begin{equation}
\sigma\w{tree}(pp\to \tilde W^0\tilde W^\pm),
\end{equation}
whose NLO-NLL values are shown in \figref{pure-ino-xs}.
Here, $\tilde W^0$ and $\tilde W^\pm$ represent the pure-winos, sharing a common tree-level mass because of the gauge symmetry.
We will first give the tree-level production cross section of
\begin{equation}
 q \overline q' \to {\tilde{\chi}^0_i} \tilde{\chi}^{\pm}_{j}\,,
\end{equation}
and then show that the production cross section in the pure-wino limit is four times larger than that in the pure-higgsino limit.
Finally, we will see that Eq.~\eqref{eq:Xsecchi21} holds.

From SU(2)$_\textrm{L}$ gauge interactions, neutralino-chargino-$W$ interactions are obtained as
\begin{equation}
\mathcal L_{\tilde \chi^0 \tilde \chi^{\mp} W^{\pm}}
 = \overline{\psi}_{\tilde{\chi}^0_i} \gamma^{\mu}\left[
(g\w L)_{ij} P\w L
+(g\w R)_{ij} P\w R
\right] \psi_{\tilde{\chi}^-_j} W_{\mu}^+
- \overline{\psi}_{\tilde{\chi}^0_i} 
\gamma^{\mu}\left[
(g\w L)_{ij}^{\ast}  P\w R
+ (g\w R)_{ij}^{\ast} P\w L
\right] \psi_{\tilde{\chi}^+_j} W_{\mu}^-
\,,
\label{eqn:chichiW-}
\end{equation}
where $g\w{L,R}$ are defined by
\beq
(g\w L)_{ij} & =- g_2  \left(  N_{i2} U_{j1}^{\ast} + \frac{1}{\sqrt{2}} N_{i3} U_{j2}^{\ast}  \right)\,,\\
(g\w R)_{ij} & =  - g_2 \left( N_{i2}^{\ast} V_{j1}  - \frac{1}{\sqrt{2}} N_{i4}^{\ast}   V_{j2}\right)\,.
\label{eqn:gLgR}
\eeq
Here, the SLHA2 notation \cite{Allanach:2008qq} is used:
in the four-component notation, the mass eigenstates are defined by
\begin{equation}
 \psi_{\tilde \chi^0_i} = \begin{pmatrix}
\tilde \chi_i^0 \\
\tilde \chi^{0\,\ast}_i
\end{pmatrix}\,,\quad
\psi_{\tilde \chi^{\pm}_j} = 
\begin{pmatrix}
\tilde \chi^{\pm}_j \\
\tilde \chi^{\mp\,\ast}_j
\end{pmatrix}\,,
\end{equation}
where
\begin{equation}
 \tilde \chi^{ 0}_i  =
 N_{ip} \pmat{- i \tilde{B} \\ -i \tilde{W}^3 \\ \tilde{H}_d^0 \\ \tilde{H}_u^0}_p\,,\quad
\tilde \chi^+_j = V_{jp} \pmat{ -i \tilde{W}^+ \\ \tilde{H}_u^+}_p\,,\quad
\tilde \chi^-_j = U_{jp} \pmat{ -i \tilde{W}^- \\ \tilde{H}_d^-}_p\,
\end{equation}
with $N_{ip}$ being $4\times4$ complex mixing matrix and $V_{ip}$ and $U_{ip}$ being $2\times2$ matrices.

Discarding the light quark mass contributions and approximating $[V\w{CKM}]_{ud}=1$, 
the squared scattering amplitude becomes 
\begin{equation}\begin{split}
&\overline{\sum\w{color}}\,\overline{\sum\w{spin}}\left|\mathcal M(u\bar d\to\neut[i]\charP[j])\right|^2
= \frac{ 2 g_2^2}{3 (s -m_W^2)^2}
\left\{\Re  \left[ (g\w L)_{ij}(g\w R)_{ij}^{\ast}\right]
  m_{{\tilde{\chi}^0_i}}m_{{\tilde{\chi}^+_j}} (p_u \cdot p_{\bar d} )
 \right.
\\
& \quad \left. + \left|(g\w L)_{ij}\right|^2   (p_{\psi_{\tilde{\chi}^0_i}}\cdot p_{\bar d})(p_{\psi_{\tilde{\chi}^+_j}}\cdot p_u) + 
 \left|(g\w R)_{ij}\right|^2  (p_{\psi_{\tilde{\chi}^0_i}}\cdot p_u)
 (p_{\psi_{\tilde{\chi}^+_j}}\cdot p_{\bar d})
 \right\}\,,
\end{split}
 \label{eq:treecrosssec}
\end{equation}
where $s \equiv (p_u + p_{\bar d})^2 \simeq 2 (p_u \cdot p_{\bar d}) $.
For the opposite-charge process 
$d \overline u \to  \tilde{\chi}^0_{i} {\tilde{\chi}^-_j}$, 
the squared amplitude is obtained by replacing Eq.~\eqref{eq:treecrosssec} as
\begin{equation}
p_{u} \to p_{d}\,,\quad
p_{\bar d} \to p_{\bar u}\,,\quad 
p_{\psi_{\tilde{\chi}^+_j}} \to p_{\psi_{\tilde{\chi}^-_j}}\,,\quad
(g\w L)_{ij} \to - (g\w R)_{ij}^{\ast}\,, \quad 
(g\w R)_{ij} \to - (g\w L)_{ij}^{\ast}\,.
\end{equation}

In the pure-wino limit, the mixing matrices satisfy
\begin{equation}
  N_{i2}=1\,, \quad V_{j1}=U_{j1}=1,
\end{equation}
which lead to
\begin{equation}
 \gL = -g_2\,, \quad \gR = -g_2\,;
\qquad
|\gL|^2=|\gR|^2=\Re(\gL\gR^*) = g_2^2\,.
\end{equation}
On the other hand, in the pure-higgsino limit, the two higgsinos form two degenerate neutralinos, and the mixing matrices become
\begin{equation}
 (N_{i3}, N_{i4})=
\begin{cases}
 \left(\frac{i}{\sqrt2},\frac{i}{\sqrt2}\right)\,,\\
 \left(\frac{-1}{\sqrt2},\frac1{\sqrt2}\right)\,,
\end{cases}
\quad V_{j2}=U_{j2}=1\,,
\end{equation}
which lead to
\begin{equation}
\gL = \frac{-N_{i3}g_2}{\sqrt2}\,, \quad \gR = \frac{N^*_{i4}g_2}{\sqrt2}\,;
\quad |\gL|^2=|\gR|^2=\Re(\gL\gR^*)=\frac{g_2^2}{4}\,.
\end{equation}
As a result, the production cross section in the pure-wino limit is four times larger than \emph{each of} the productions in the pure-higgsino limit.
This property holds well at the NLO-NLL level, as shown in \figref{pure-ino-xs}.

Next, let us define 
\begin{equation}
  (c\w{LL})_{ij} = \frac{|(\gL)_{ij}|^2}{g_2^2}\,,
\quad
  (c\w{RR})_{ij} = \frac{|(\gR)_{ij}|^2}{g_2^2}\,,
  \quad
  (c\w{LR})_{ij} = \frac{\Re[(\gL)_{ij}(\gR)_{ij}^\ast]}{g_2^2}\,,\label{eq:c}
\end{equation}
and
\begin{equation}\label{eq:RNCdef}
   R^{\mathrm{NC}}_{ij} = \mathop{\mathrm{mean}}\left[(c\w{LL})_{ij},(c\w{RR})_{ij},(c\w{LR})_{ij}\right]\,.
\end{equation}
In the pure-wino limit, $c\w{LL}=c\w{RR}=c\w{LR}=1$, while in the pure-higgsino limit they are $1/4$.
From Eq.~\eqref{eq:treecrosssec} it is found that, if $c\w{LL}=c\w{RR}=c\w{LR}$, the cross section is given by
\begin{equation}
  {\sigma_{\text{tree}}(pp\to\neut[i]\charPM[j])} = R_{ij}^{\mathrm{NC}}\times
  {\sigma_{\text{tree}}(pp\to\neut[i]\charPM[j])|\w{wino}}\,,
\end{equation}
and that the kinematic distribution of the produced electroweakinos are the same as its wino-like limit at the leading order.

\begin{figure}[t]
 \centering \fourfigs{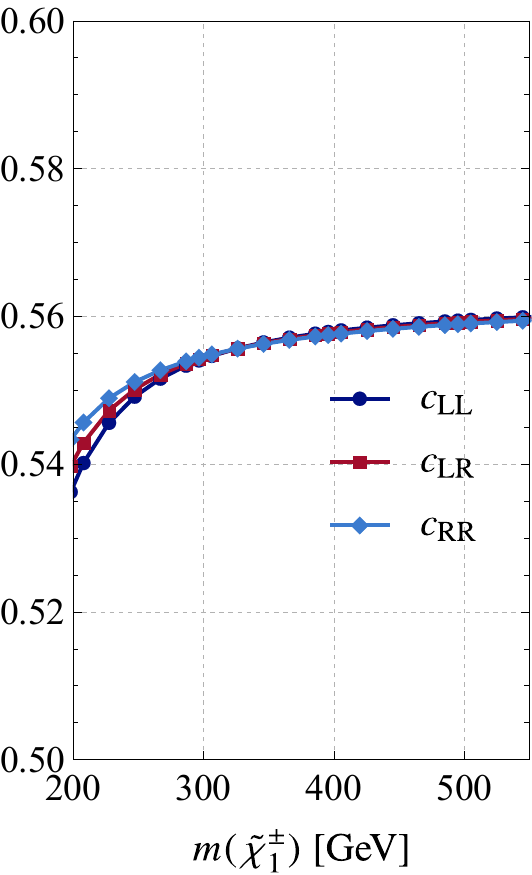}
\caption{The numerical values of $c\w{LL,LR,RR}$ defined in Eq.~\eqref{eq:c} for $\neut[2]\charPM[1]$ production in our models (A)--(D) from left to right. They are independent of the slepton masses because we calculate the electroweakino masses and mixings at the tree-level.}
\label{fig:n2c1cfactor}
\end{figure}

This relation $c\w{LL}= c\w{RR}= c\w{LR}$ is found to hold well in our model points, especially for $(i,j)=(2,1)$, as is displayed in \figref{n2c1cfactor}.
Therefore, because $\neut[2]$ and $\charPM[1]$ are degenerate in our model points, the production cross section is approximately given by
\begin{equation}
 \sigma\w{tree}(pp\to\neut[2]\charPM[1])\approx R^{\mathrm{NC}}_{21}\times\sigma\w{tree}(pp\to\tilde W^0\tilde W^\pm)\,.
\end{equation}
We utilized this relation in Eq.~\eqref{eq:Xsecchi21}.

\section{List of LHC results in our analysis}
\label{app:LHC_analysis_list}

This appendix summarizes the LHC Run~2 results ($\sqrt s=13\TeV$) compiled in our analysis.

The exclusion regions in \figref{summary} are based on
\begin{itemize}
 \item Ref.~\cite{Aad:2019vnb}      by ATLAS collaboration (SLSL, $139\ifb$),            
 \item Ref.~\cite{Aad:2019vvf}      by ATLAS collaboration (CC/WW and NC/HW, $139\ifb$), 
 \item Ref.~\cite{Sirunyan:2017lae} by CMS   collaboration (NC/3L, $35.9\ifb$),          
 \item Ref.~\cite{Aaboud:2018jiw}   by ATLAS collaboration (NC/3L, $36.1\ifb$).          
\end{itemize}
Upper bounds for the following references are also considered in \figref{n2c1summary}:
\begin{itemize}
 \item Ref.~\cite{Sirunyan:2018ubx} by CMS   collaboration (NC/HW and NC/ZW, $35.9\ifb$), 
 \item Ref.~\cite{Aaboud:2018sua}   by ATLAS collaboration (NC/ZW, $36.1\ifb$),           
 \item Ref.~\cite{Aad:2019vvi}      by ATLAS collaboration (NC/ZW, $139\ifb$).            
\end{itemize}
In addition, we have checked the following works, which are possibly relevant for our model points:
\begin{itemize}
 \item Ref.~\cite{Sirunyan:2018nwe} by CMS   collaboration, $35.9\ifb$, 
 \item Ref.~\cite{Sirunyan:2018lul} by CMS   collaboration, $35.9\ifb$, 
 \item Ref.~\cite{Sirunyan:2019iwo} by CMS   collaboration, $77.5\ifb$, 
 \item Ref.~\cite{Sirunyan:2017qaj} by CMS   collaboration, $35.9\ifb$, 
 \item Ref.~\cite{Aad:2019qnd}      by ATLAS collaboration, $139\ifb$,  
 \item Ref.~\cite{Sirunyan:2018iwl} by CMS   collaboration, $35.9\ifb$, 
 \item Ref.~\cite{Sirunyan:2019zfq} by CMS   collaboration, $35.9\ifb$, 
 \item Ref.~\cite{Sirunyan:2018vig} by CMS   collaboration, $35.9\ifb$, 
 \item Ref.~\cite{CMS-PAS-SUS-17-002} by CMS collaboration, $35.9\ifb$, 
 \item Ref.~\cite{CMS:2019eln}      by CMS   collaboration, $77.2\ifb$, 
 \item Ref.~\cite{Sirunyan:2019mlu} by CMS   collaboration, $77.2\ifb$, 
 \item Ref.~\cite{Aad:2019byo}      by ATLAS collaboration, $139\ifb$.  
\end{itemize}

\clearpage

\section{2021 spring update}
\label{sec:update}
\label{sec:update2021spring}


\noindent Recently, the muon $g-2$ collaboration presented a new result on the measurement of $\amu$ at the Fermilab~\cite{g-2Seminar20210407}:
\begin{equation}
 \amu[\text{FNAL; 2021 Apr.}] =
\left(11\,659\,204.0 \pm 5.4\right) \times 10^{-10}\,.
\end{equation}
Together with the BNL result, Eq.~\eqref{eq:amuBNL}, the state-of-art value of the $\amu$ measurements is given by
\begin{equation}
 \amu[\text{BNL+FNAL}] =
\left(11\,659\,206.1 \pm 4.1 \right) \times 10^{-10}\,.
\end{equation}
This value corresponds to $4.2\,\sigma$ level discrepancy when compared to the SM prediction given in Ref.~\cite{Aoyama:2020ynm}\footnote{%
 Several new lattice results for the SM hadronic vacuum polarization contribution~\cite{Borsanyi:2020mff,Lehner:2020crt} are not included in this combination.}:
\begin{align}
 &\amu[\text{SM}] = \left( 11\,659\,181.0 \pm 4.3 \right) \times 10^{-10}\,;\\
 &\Delta a_{\mu}  = \left( 25.1\pm 5.9\right) \times 10^{-10}\,.
\end{align}
Here we present an updated version of Fig.~\ref{fig:summary} based on this new value of $\Delta\amu$.

We also update LHC constraints by including new results from the ATLAS and CMS collaborations.
There are five new results that are relevant for our parameter space.
We analyzed two of them,
\begin{itemize}
 \item Ref.~\cite{Aad:2020qnn}         by ATLAS collaboration (NC/HW, $139\ifb$),           
 \item Ref.~\cite{Sirunyan:2020eab}    by CMS collaboration (SLSL and NC/ZW, $139\ifb$),    
\end{itemize}
which are prepared as publication, by the procedures described in Sec.~\ref{sec:LHC}.
It is found that our parameter space is partially excluded by the SLSL analysis based on Ref.~\cite{Sirunyan:2020eab} as shown in Fig.~\ref{fig:summary2021spring} (blue dash-dotted region labeled by ``SLSL/C'').
The ``SLSL/A'' blue-filled region has been excluded by the ATLAS counterpart \cite{Aad:2019vnb} corresponding to Fig.~\ref{fig:summary}.
Meanwhile, we found that no additional region is excluded by NC/HW and NC/ZW processes, which is partially because of the small branching ratio of $\neut[2]\to Z\neut[1]$.

The other three works~\cite{ATLAS-CONF-2020-015,CMS-PAS-SUS-19-012,CMS-PAS-SUS-20-003} are preliminary conference papers and not included in this update because numerical data have not been made public.
According to our estimation, they provide additional constraints based on the NC/3L and NC/HW signatures.
The new NC/3L result~\cite{CMS-PAS-SUS-19-012} will confirm that our parameter spaces with $m_{\tilde{\mu}\w L} < m_{\tilde{\chi}^{\pm}_1}$ are strongly disfavored, while the new NC/HW result~\cite{CMS-PAS-SUS-20-003} will exclude more region with $m_{\tilde{\mu}\w L} > m_{\tilde{\chi}^{\pm}_1}$, where the impact depends on the LSP mass.


In summary, we provide Fig.~\ref{fig:summary2021spring}, which supersedes Fig.~\ref{fig:summary}.
It incorporates the new $\amu$ measurements~\cite{g-2Seminar20210407}, new theory combination~\cite{Aoyama:2020ynm}, and new results from the LHC~\cite{Aad:2020qnn,Sirunyan:2020eab}.
We find that, as far as considering the parameter space in Fig.~\ref{fig:summary2021spring}, models with $m_{\tilde{\mu}\w L} < m_{\tilde{\chi}^{\pm}_1}$ are strongly disfavored as a solution to the muon $g-2$ anomaly.
Meanwhile, for models with $m_{\tilde{\mu}\w L} > m_{\tilde{\chi}^{\pm}_1}$, LHC constraints are not critical yet; those models may explain the muon $g-2$ anomaly and are to be searched for at the future LHC runs.

\begin{figure}[p]
 \centering
 \renewcommand\thesubfigure{\Alph{subfigure}}
  \begin{subfigure}[b]{0.49\textwidth}
 \includegraphics[width=\textwidth]{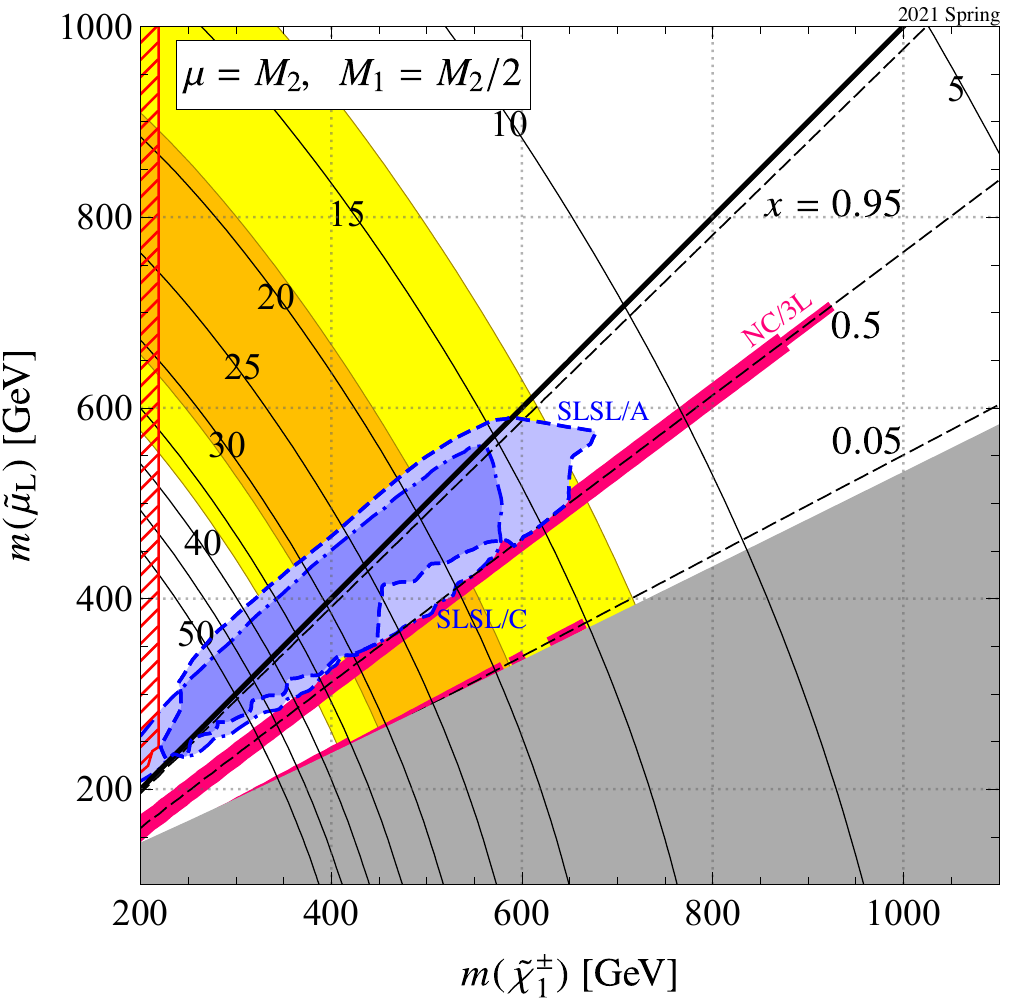}
\caption{$\mu = M_2$, $M_1 = M_2/2$}
 \vspace{.2cm}
 \end{subfigure}
   \begin{subfigure}[b]{0.49\textwidth}
 \includegraphics[width=\textwidth]{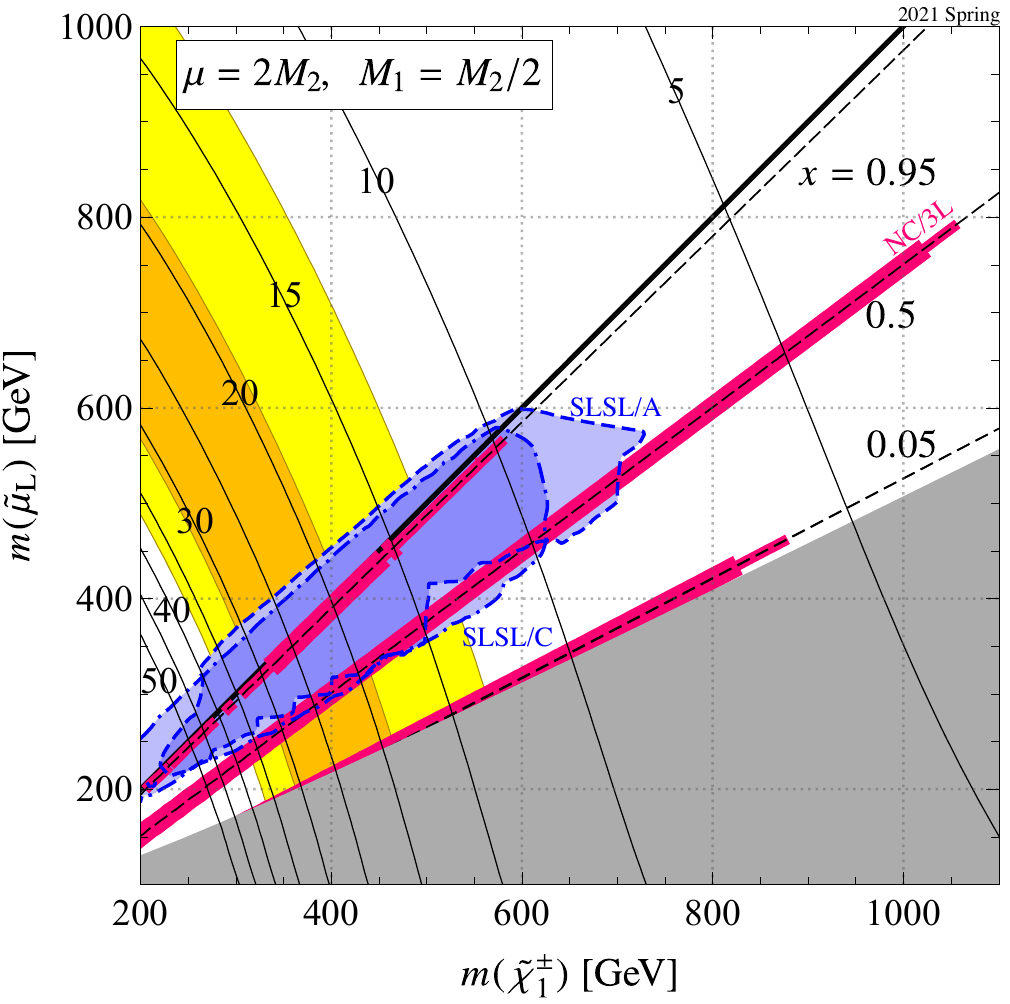}
\caption{$\mu = 2 M_2$, $M_1 = M_2/2$}
 \vspace{.2cm}
 \end{subfigure}
   \begin{subfigure}[b]{0.49\textwidth}
 \includegraphics[width=\textwidth]{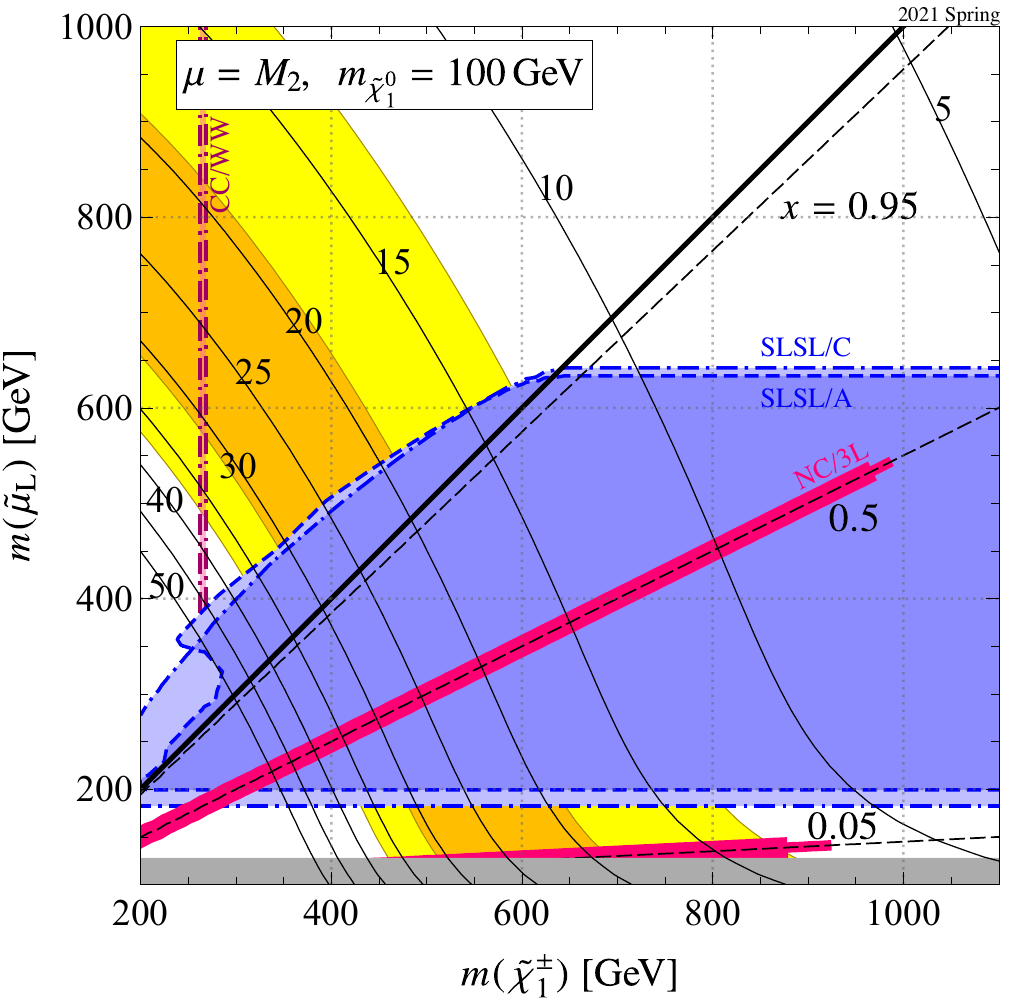}
\caption{$\mu = M_2$, $m_{\tilde{\chi}^0_1}=100$\,GeV}
 \end{subfigure}
   \begin{subfigure}[b]{0.49\textwidth}
 \includegraphics[width=\textwidth]{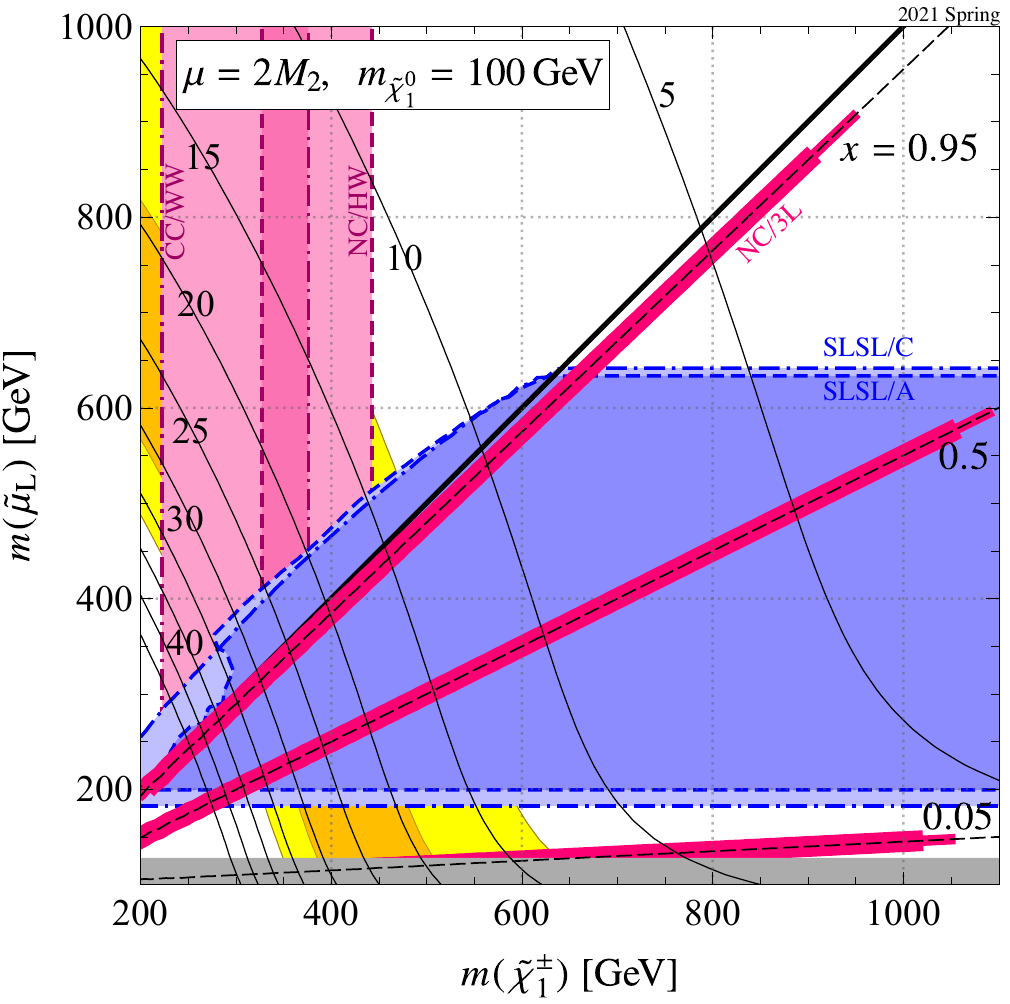}
\caption{$\mu = 2 M_2$, $m_{\tilde{\chi}^0_1}=100$\,GeV}
 \end{subfigure}
  \caption{\label{fig:summary2021spring}%
  The 2021 Spring Update of Fig.~\ref{fig:summary}, a summary of chargino-dominated SUSY scenario for the muon $g-2$ anomaly.
  The black contours show $\amu[SUSY]\times10^{10}$ but only up to $50$.
  In the orange-filled (yellow-filled) regions, $\amu[SUSY]=(25.1\pm5.9)\times10^{-10}$ is satisfied at the $1\sigma$ ($2\sigma$) level.
  The thick black line corresponds to $m_{\tilde{\mu}\w L} = m_{\charPM[1]}$.
  The gray-filled region, where the LSP is $\tilde\nu$, and the red-hatched region in (A), which corresponds to a compressed spectrum, are not studied.
  The LHC constraint from the CC/WW (NC/HW) analysis is shown by the red-filled regions with the dash-dotted (dashed) boundaries.
  The blue-filled regions are excluded by the SLSL analysis: the blue dashed lines denote constraint from the ATLAS experiment (SLSL/A), while the blue dash-dotted lines are the newly-included constraint from the CMS experiment (SLSL/C).
  The constraints from the NC/3L analysis are investigated on the model points with $x=0.05$, $0.5$, and $0.95$ (see Eq.~\eqref{eq:x}), where the exclusion ranges are shown by the magenta lines.
  }
\end{figure}

\bibliography{ref}

\end{document}
